\documentclass[nofootinbib,superscriptaddress, aps, prd, 10pt, amsmath, amssymb, bibnotes,
altaffilletter, twocolumn, floatfix]{revtex4-2}

\usepackage{hyperref}
\usepackage[T1]{fontenc}
\usepackage{lineno}
\usepackage{xcolor}
\usepackage{graphicx}
\usepackage{dcolumn}
\usepackage{bm}
\usepackage{upgreek}
\usepackage{tabularx}

\usepackage{lipsum}  

\hypersetup{colorlinks=true, citecolor=blue, urlcolor=blue, linkcolor=blue}

\definecolor{mycolor}{rgb}{0.3, 0.2, 0.9}
\begin{document}

\title{Athermal phonon collection efficiency in diamond crystals for low mass dark matter detection }

\newcommand{\lanl}{Los Alamos National Laboratory, Los Alamos, NM 87545, USA}
\newcommand{\llnl}{Lawrence Livermore National Laboratory, Livermore, CA 94550, USA}
\newcommand{\ohio}{Department of Physics, Ohio State University, Columbus, OH 43210, USA}
\newcommand{\slac}{SLAC National Accelerator Laboratory, Menlo Park, CA 94025, USA}
\newcommand{\unm}{Department of Physics and Astronomy, University of New Mexico,  Albuquerque, NM 87131, USA}

\author{I.~Kim}~\email{kim124@llnl.gov}\affiliation{\llnl} 
\author{N. A. Kurinsky}\affiliation{\slac}
\author{H.~Kagan}\affiliation{\ohio}
\author{S. T. P. Boyd}\affiliation{\unm}
\author{G. B. Kim}\affiliation{\llnl}

\noaffiliation

\date{\today}

\begin{abstract}

We explored the efficacy of lab-grown diamonds as potential target materials for the direct detection of sub-GeV dark matter~(DM) using metallic magnetic calorimeters~(MMCs). 
Diamond, with its excellent phononic properties and the low atomic mass of the constituent carbon, can play a crucial role in detecting low-mass dark matter particles. 
The relatively long electron-hole pair lifetime inside the crystal may provide discrimination power between the DM-induced nuclear recoil events and the background-induced electron recoil events. 
Utilizing the fast response times of the MMCs and their unique geometric versatility, we deployed a novel methodology for quantifying phonon dynamics inside diamond crystals. 
We demonstrated that lab-grown diamond crystals fabricated via the chemical vapor deposition~(CVD) technique can satisfy the stringent quality requirements for sub-GeV dark matter searches. 
The high-quality polycrystalline CVD diamond showed a superior athermal phonon collection efficiency compared to that of the reference sapphire crystal, and achieved energy resolution 62.7~eV at the 8.05~keV copper fluorescence line. 
With this energy resolution, we explored the low-energy range below 100~eV and confirmed the existence of so-called low-energy excess~(LEE) reported by multiple cryogenic experiments. 
\end{abstract}

\maketitle

\section{Introduction}
Even with the growing sizes of the large-scale experiments, the dark matter~(DM), which is thought to comprise approximately 85\% of the matter in the universe, has so far evaded detection in terrestrial experiments~\cite{Clowe_2006_DM,liu2017current,Undagoitia_2015,choi2015skk,agnese2016cdms,armengaud2016edelweiss,arnaud2020edelweiss,conrad2017indirect,Abbasi2022icecube,pandax4t2021dm,xenonnt2022dm}. 
The DM particle could have evaded direct detection if their interaction with the standard model particle produce energy deposits lower than the threshold of the conventional detectors. 
As a result, searches for sub-GeV ``light mass'' dark matter have been receiving increasing attention in recent decades~\cite{battaglieri2017us,boveia2022snowmass}. 

So far, the DM-nucleus cross-section has been poorly constrained in the sub-GeV DM mass~($m_\chi$) range~\cite{particle2022reviewDM}, primarily because the recoil energy from nuclear scattering is typically below the threshold of conventional detectors.
Many significant efforts are going on to overcome this problem, either by scanning new types of interactions that were not previously accounted for, or by developing detection techniques to lower the nuclear recoil threshold.
Scanning new types of interactions includes searching for DM-electron interaction~\cite{essig2012subgev,graham2012subgev,essig2016direct}, scanning for signatures from exotic dark matter models~\cite{pospelov2008superwimp,gerda2020superwimp,arnquist2022exotic}, or exploiting the theoretical ``Migdal effect'' where a slow nuclear recoil is thought to have a substantial probability of shaking off electrons from the nucleus, thus creating a secondary ionization signal~\cite{vergados2005migdal,bernabei2007migdal,ibe2018migdal,essig2020migdal}.  
On the other hand, developing lower nuclear recoil threshold includes developing entirely novel detector concepts~\cite{tiffenberg2017sccd,fichet2018quantumdm,guo2013superfluid,caputo2020superfluid,maris2017evaporation,hochberg2017grephene,kim2020detection,hochberg2019nanowire,carter2017calorimetric,essig2017chemical} and improving the sensitivity of existing detectors by using advanced target material~\cite{hochberg2016superconductor,hochberg2018dirac,hochberg2016detecting,kurinsky2019diamond}.

Diamond, in particular, has been the subject of extensive studies as an ideal candidate for the low-mass DM searches in the past few years~\cite{kurinsky2019diamond,canonica2020diamond,marshall2021diamond,abdelhameed2022diamond}, owing to its high Debye temperature~(2220~K), high isotopic purity, excellent phonon properties, radiation hardness, good semiconductor properties, ease of handling, and the low atomic mass of carbon. 
The properties of diamond crystals as the DM detector target material are well summarized in Ref.~\cite{kurinsky2019diamond}.
So far, building detectors with bulk diamond crystals has been out of reach because of the high cost of the material. 
However, recent advancements in the chemical vapor deposition~(CVD) techniques has enabled the production of high-quality ``detector grade'' diamond crystals at an affordable cost. 
Building a bulk detector with diamond target crystals is within realistic reach of the next generation sub-GeV dark matter detection experiments.

Crystal-based rare event searches typically measure incident particles' energy by coupling sensors to large crystal particle absorbers. 
For more than a decade, quantum sensors operated at low temperatures, such as transition edge sensors~(TES) and metallic magnetic calorimeters~(MMCs) have been extensively used in various experiments exploring rare processes, for their high energy resolutions~\cite{kempf2018physics}. 
In particular, MMC phonon sensors feature sub-microsecond response times and high energy resolution~\cite{Fleischmann2005}, thereby providing a capability of discriminating the DM-induced nuclear recoil~(NR) signals from the electron recoil~(ER) signals from background radiation~\cite{boyd2023development}, a crucial aspect in the pursuit of a background-free dark matter search.  
Furthermore, MMC-based particle detectors offer high versatility by enabling the fabrication of the phonon absorber in a variety of shapes, independent of the phonon sensor's configuration~\cite{amore2017novel}. 
This unique aspect is enabled by the spatial separation between the phonon absorber and the MMC sensor within the detector setup. 
Such a design affords us the capability to fabricate the phonon absorbers in various geometries and allows us to test multiple crystals in controlled and reliable ways, without impinging upon the functionality or integration of the phonon sensor. 
The flexibility in design, allowing for various geometrical configurations of the absorbers, enables further innovations in detector design aimed at reducing the energy threshold even more. 

In this article, we introduce a novel method for analyzing the evolution of phonons within a crystal upon collision of energetic particles. 
Utilizing MMC quantum sensors with sub-microsecond response times, we are testing phonon collection efficiencies and quasiparticle lifetimes of various crystals to test their efficacy as potential target materials for the sub-GeV dark matter detectors. 
In this work, we demonstrate that lab-grown diamond crystals fabricated via the chemical vapor deposition technique can meet the stringent quality requirements for low-threshold  measurements.
We also present the first low-energy measurements obtained from an 
MMC-based detector with diamond target, achieving energy threshold below 100~eV. 
Our results highlight the potential of integrating high-quality lab-grown diamond with MMC quantum sensors to significantly enhance the dark matter detection sensitivity.

\begin{figure}[t]
    \centering
    \includegraphics[width=\columnwidth]{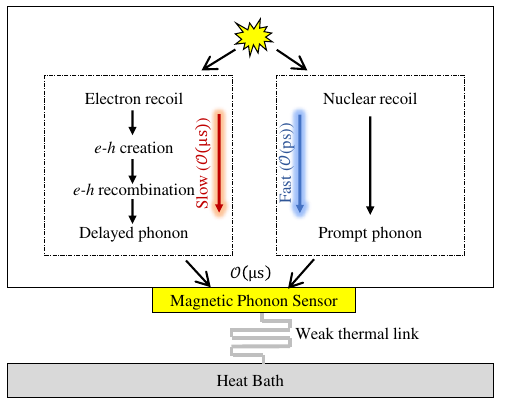}
    \caption{Schematic diagram of phonon creation mechanism in crystal, and phonon collection scheme. See text for discussions.}
    \label{fig:psdExplanation}
\end{figure}

\section{Concept}

Crystal-based phonon-sensing particle detectors are typically composed of three main components: a bulk crystal absorber where particles interact, a thin gold film that serves as a phonon collector, and a phonon sensor that measures the total energy deposited by the particle interaction. 
Incident particles can deposit energy in the crystal through either an NR or an ER process.

Although many aspects of electron and nuclear recoil in diamond are well understood, the exact energy partitioning and the lifetimes of charge carriers can vary with temperature, and the interaction processes at mK temperatures are yet to be fully characterized. One of the primary motivations for phonon-based sensing in diamond detectors, such as this experiment, is to investigate these low temperature dynamics in greater detail.

In electron recoil events~(from background radiation), the incident particle primarily deposits its energy into electronic excitations in diamond, creating electron-hole~($e-h$)~pairs in diamond.
The $e-h$ pairs recombine on a time scale ranging from a few nanoseconds to milliseconds, depending on the material and the quality of the crystals~\cite{stevenson1955carrier,isberg2002high,bi2016charge,zhang2017perovskite,gradwohl2021carrier}.
The $e-h$ pair recombination process creates ballistic athermal phonons that are collected in the metallic phonon collector. 
The athermal phonon signal rise time is thus set by the pair recombination time. 
Athermal phonons may also thermalize into the phonon bath before they can be collected, creating a secondary phonon signal.

In contrast, for nuclear recoil events~(expected to be from dark matter candidates or neutrons) at recoil energies below $\approx$10~keV, more than 50\% of the recoil energy is immediately deposited as prompt phonons~\cite{deboer2007radiation}. 
This difference in the timescale of phonon creation is key to discriminating between ER and NR, especially in applications such as dark matter searches and other rare process searches where background rejection is critical.

Energy loss to de-excitation photons in $e-h$ pair recombination is expected to be negligible. 
In high-purity CVD diamond, the radiative exciton lifetime is measured to be 700–800~ns, and was temperature-independent between 100~K and 295~K~\cite{kozak2013optical}. 
The competing nonradiative Auger recombination occurs on a $\approx$1~ns timescale at 15~K, and is much faster than radiative recombination at this temperature~\cite{shimano2002formation}. 
Further, excitons undergo internal thermalization within $\approx$5~ps, meaning that excess energy is quickly dissipated into the phonon bath before recombination occurs~\cite{shimano2002formation,akimoto2014high}.

In semiconductor diamond particle detector setups, where collected charge is the signal, it is found that the average energy required to form one $e-h$ pair is $\approx$13~eV, while the band gap is 5.47~eV. 
Thus, roughly 40\% of the recoil energy partitioned to $e-h$ pair excitation ultimately appears as free charges~\cite{ziaja2005cascades,kurinsky2019diamond}, while the remainder is deposited into the phonon bath and become undetectable in semiconductor detectors. 
By sensing phonons directly, the present experiment can sense the full recoil energy.
Further, it is insensitive to the statistics of charge collection and their impact on energy resolution.

A similar pulse shape discrimination~(PSD) concept has been applied to detectors with scintillating crystals~\cite{amore2017novel,kim2020self}, exploiting the very slow scintillation decay time at mK temperatures~\cite{amore2015scintillation}, and demonstrated good ER/NR separation. 
However, for semiconductor detectors like diamond, where the electron-hole recombination time can be much faster, on the order of microseconds, a faster sensor response time is required to achieve similar discrimination power. 
Figure~\ref{fig:psdExplanation} shows the detector concept for the MAGNETO-DM~\cite{magnetodm2023} dark matter search, visualizing the concept of PSD in semiconductor crystals. 
Fast phonon sensors of MAGNETO-DM with sub-microsecond response times will enable the discrimination between ER and NR signals. 

The ability to discriminate between NR and ER signals depends mainly on three timescales: the $e-h$ pair recombination time, the phonon collection time in the crystal, and the detector response time. 
With the sub-microsecond response times the MMC sensors provide, the relatively long  $e-h$~  pair lifetimes, especially in a high-quality CVD diamond where  lifetimes can exceed 2~$\upmu$s ~\cite{isberg2002high}, can provide measurable pulse shape differences between the NR and the ER signals.
Additionally, diamond's high sound velocity enhances phonon collection efficiency, improving energy resolution and signal discrimination. 

The fast response time of MMCs and their high versatility also allow us to test the phonon evolution  inside the target crystal under various conditions. 
If an external particle hits the gold phonon collector directly, the particle energy is quickly transferred to the electron system of the metal, and all of its energy can be read by the sensor. 
The pulse shape of these signals provides a benchmark for how the sensor responds to a single particle energy deposition. 
While electron recoils directly energize conduction electrons, nuclear recoils excite energetic phonons that subsequently energize electrons. 
However, this time scale is in $\mathcal{O}$(100~fs) in a thin gold film~\cite{sun1994femtosecond,fann1992electron} and does not create a measurable difference in the benchmark pulse shape, even with the fast MMC sensors. 
Hence, the \textit{detector response} pulse shape is always identical regardless of the type of particle, whether it is $\alpha$, $\beta/\gamma$, or even a quasiparticle like a phonon that deposits the energy into the absorber. 
In contrast, in case of the crystal hit events (either ER or NR) that creates a large number of acoustic phonon inside the crystal, each phonon arrival will create a single detector response.
These athermal acoustic phonons can be considered to first order as ballistic quasiparticles travelling inside the crystal, which can be reflected from boundaries, transmitted to the gold phonon collector, lost to the heat bath via the weak thermal link, or down-converted to thermal phonons. 
A small proportion of the athermal phonons transferred to the gold absorber may also reflect back to the crystal, but the probability is low due to the fast thermalization time in gold film. 
The signal induced by the particle absorbed in the crystal therefore appears as the convolution of the phonon evolution in the crystal and the detector response~\cite{probst1995model}. 
Hence, we can decouple the pure phonon evolution from the detector properties via deconvolving the pulse shape of the crystal hit events from the detector response. 

Only the phonons arriving at the phonon collector can be read by the MMC sensor, causing some inefficiency that is heavily dependent not only on the crystal geometry but also on the properties such as the quality of the crystal lattice. 
Hence, quantifying the phonon evolution inside the crystal is one of the most direct ways to study the properties of crystal samples. 

\begin{figure}[t]
    \centering
    \includegraphics[width=\columnwidth]{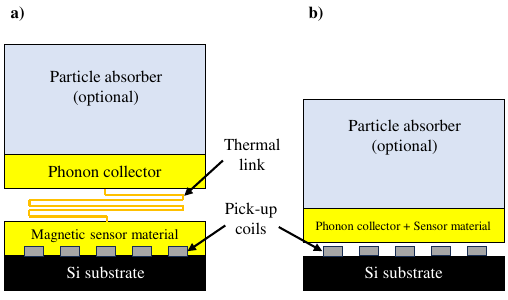}
    \caption{Conceptual designs of the MMC-based particle detector, equipped with an optional particle absorber crystal. (a) conventional layout with thermal link connecting the phonon collector and the paramagnetic sensor material. The thermal link, gold wire bond, is typically $\mathcal{O}$(mm) long. (b) next generation direct-contact setup. The paramagnetic sensor material and the phonon collector form a single body. The conventional setup was used in this work, while the direct-contact setup will be used in future.}
    \label{fig:UF}
 \end{figure}
 
\section{Experimental Setup}

To test the concept of an MMC-based diamond detector, we used two small diamond crystals as test setups. 
In the first setup, we used a ($10\times10\times0.5$)~mm$^3$ polycrystalline~CVD~(pCVD) diamond from II-VI~\cite{iivi2023}. 
An $^{241}$Am source was placed over the diamond crystal with copper $\alpha$ blocker for energy calibration. 
The expected characteristic energy lines include 59.54~keV, 26.34~keV and 33.20~keV $\gamma$-rays from the nuclear decay, and the characteristic 13.9~keV L$_{\alpha1}$, 17.8~keV L$_{\beta1}$ and 20.8~keV L$_{\gamma1}$ x-ray lines from $^{237}$Np. 
The 8.05~keV and 8.91~keV x-ray fluorescence lines from the copper served as additional calibration peaks in the low-energy region.
The total event rate was controlled to be approximately 10~Hz to allow enough time for the detector to fully thermalize back to the base temperature before the subsequent event triggers. 
We installed MMC sensors fabricated by the Korea Research Institute of Standards and Science~(KRISS) and Institute for Basic Science~(IBS) with a Magnicon C6XS1 SQUID to measure the phonons in the diamond crystal. 
A thin gold film of 3~mm~$\times$~1.5~mm~$\times$~300~nm with a 2~nm Cr adhesion layer was evaporated directly onto the surface of the crystal to serve as the phonon collector. 
The thermal connection between the gold phonon collector and the MMC sensor was made by gold wire bonds in this measurement, as illustrated in Fig.~\ref{fig:UF} a).  
This connection is to be improved in future by advancing to a direct mechanical contact~\cite{boyd2023development} to enhance the speed of the heat flow, as shown in Fig.~\ref{fig:UF} b). 
Removal of the thermal link, which is typically $\mathcal{O}$(mm), enhances the heat flow from the absorber to the sensor material, thereby increasing the speed of the signal rise.

\begin{table}[t]
    \centering
    \begin{tabular}{|c|c|c|c|}
         \hline
         & 8.05~keV & 26.34~keV & 59.54~keV \\
         \hline
        0.5mm diamond~(pCVD) & 52.4& 1.6 & 0.1\\
        0.3mm diamond~(sCVD) & 35.9& 1.0 & 0.06\\
        0.5mm sapphire & 99.8 & 15.5& 1.2\\
        \hline
        8$\upmu$m MMC sensor & 95.5& 44.5& 5.9\\
        300nm gold absorber & 11.0& 2.2& 0.2\\
        25$\upmu$m diameter gold wire & 100.0& 84.1& 17.4\\
         \hline
    \end{tabular}
    \caption{Photoelectric absorption probability for perpendicularly incident $\gamma$-radiations at each part of the detector setup. Quantities are in percentage.}
    \label{tab:interactionProbability}
\end{table}

In the second setup, the pCVD crystal was replaced with a ($5\times5\times$0.3)~mm$^3$ single-crystal CVD~(sCVD) diamond from Element~Six~\cite{e6}.
We also tested a similar detector setup with a ($5\times5\times$0.5)~mm$^3$ sapphire crystal as the target material as the control setup. 
In Table.~\ref{tab:interactionProbability}, the absorption probabilities for the three most prominent $\gamma-$photopeaks are summarized for the diamond and sapphire crystals. 
Compton scattering becomes dominant over 22~keV and 45~keV for diamond and sapphire.
Only a fraction of higher energy photons would deposit the full energy to the crystal targets, especially for the diamond. 
By examining the relative heights of the $\gamma-$peaks in combination with the pulse shape analaysis, one can identify at which volume an event is originated from. 

To ensure a complete data acquisition, a continuous stream of data was collected at 1MHz sampling rate using the 14-bit NI PXIe-5172 digitizer~\cite{nationalinstrument}.
An SR-560 low noise voltage preamplifier~\cite{srs} was installed before the digitizer to enhance the signal-to-noise ratio. 
The dynamic range of the device was set to capture the 59.4~keV $\gamma$ pulses.
6dB/oct rolloff highpass filter was applied to mitigate low-frequency noise induced by acoustic vibration. 
The filtering condition was optimized for each dataset independently to address changes in noise conditions. 
As the thermal link via wire bonding has slowed the pulses down to have a rise time of a few tens of $\upmu$s, the high-frequency cutoff at 1~MHz did not affect the analysis. 
While discriminating the NR and the ER events is beyond the scope of this article, a higher bandwidth may be required in future work for the particle discrimination.
For the sapphire control setup the sampling rate was limited to 0.5~MHz. 

Two sets of measurements were made separately for each diamond crystal. 
The first sets of measurement focused on achieving high energy resolution dataset with heavy high-pass filtering, to precisely study the energy deposition by incoming particle to various volumes comprising the detector. 
In the second sets of measurement, we used minimal high-pass filtering for the pulse shape analysis. 
Measurement times were longer for the sCVD, as its volume was 6.7 times smaller than the pCVD. 
The details of the two measurements are summarized in Table~\ref{tab:crystals}.
% Run 80: sCVD
% Run 88: sCVD
% Run 89: pCVD 
% Run 97: pCVD

% at 100~Hz 

\begin{table*}[t]
    \centering
    % \footnotesize
    \begin{tabular}{|c|c|c|c|c||c|}
    \hline
     & Setup 1-1 & Setup 1-2 & Setup 2-1&Setup 2-2& control \\
    \hline
    Crystal & poly CVD & poly CVD & single CVD  &single CVD& sapphire\\
    Dimension (mm$^3$) & $10\times 10\times 0.5$ & $10\times 10\times 0.5$ & $5\times 5\times 0.3$  &$5\times 5\times 0.3$  & $5\times 5\times 0.5$ \\
 Dataset& High resolution& Minimal filtering& High resolution& Minimal filtering&Control\\
    Temperature (mK) & <10 & <10 & <10  &<10  & <10\\
    Collection time (hours) & 3.96& 3.47& 8.92&15.3& 4.17 \\
    Sampling rate (MHz) & 2& 1& 1  &1& 0.5\\
    Bandwidth~(kHz)& [ 0.1, 300 ]& [ 0.1, 500 ]& [ 0.1, 500 ] &[ 0.1, 500 ] & [ 0.1, 250 ]\\
    Filtering frequency~(kHz)& [ 0.5, - ]& [ 0.1, - ]& [ 1, - ]&[0.1, -]&[ 0.1, - ]\\
    \hline
    \end{tabular}
    \caption{Measurement details of the two diamond setups and the sapphire data control setup. }
    \label{tab:crystals}
\end{table*}

\section{Analysis}

\subsection{Pulse shape discrimination}

The MMCs can read out phonons from any part of the setup if their energy can be transmitted to the sensor. 
The detector setup used for this experiment is composed of: the diamond crystal, the gold phonon collector, the Au:Er MMC sensor material, and the silicon substrate on which the MMC is deposited. 
An external particle can deposit energy to any of these materials. 
The heat flows sequentially in the order of the crystal, the gold phonon collector, the weak thermal link, the MMC sensor, and the substrate where it flows out to the bath.
Hence, the pulse rise time is the fastest in the sensor and the slowest in the crystal.
On the other hand, the decay times of the events occurring in the collector and the sensor will be the same, as they are fully thermalized within $\approx 100~\upmu$s. 
For crystal events, however, the thermal phonon contribution produces a continuous heat input to the phonon collector~($\mathcal{O}$(ms)), slowing down the decay of the pulses. 
This capability of the pulse shape discrimination~(PSD) enables the unique study about the phonon evolution inside the crystalline absorber.

\begin{figure}
    \centering
    \includegraphics[width=\columnwidth]{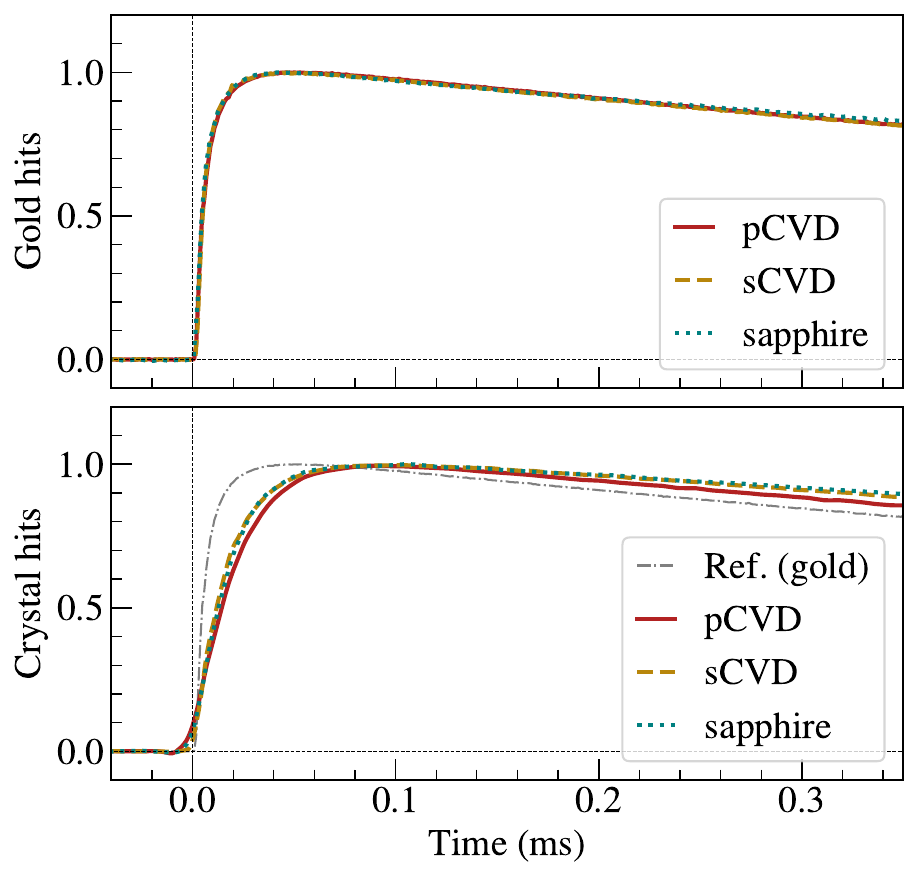}
    \caption{Comparison of the 59.54~keV events in three different datasets. (top) gold-hit events.  Good agreement between  waveforms demonstrates the identical detector responses in three different measurements. (bottom) crystal-hit events. pCVD crystal shows lower thermal component.}
    \label{fig:rising}
\end{figure}

While the waveforms of the crystal-hit events are affected by the phonon dynamics inside the crystal, the waveforms of the gold-hit events are invariant upon change of crystals. 
As the geometry of the detector setup was controlled to be identical in all measurements aside from the target crystal, the detector responses in the three measurements would also be identical, as demonstrated by the 59.54~keV gold-hit waveforms in Fig.~\ref{fig:rising}. 
The three waveforms from minimal filtering datasets showed extremely good agreement with one another. 
This serves as an important control when comparing phonon dynamics inside different crystals. 
Any difference in crystal-hit waveforms must originate from the difference in the target crystal.

We applied two trapezoid filters~\cite{jordanov1994digital} with different shaping parameters identically to the high-resolution datasets and the sapphire control set to extract the energy of the incident particle and the pulse shape information.
The long trapezoid with the shaping time of $\tau_\textrm{shape}=100~\upmu$s and the flattop time of $\tau_\textrm{flat}=60~\upmu$s was used to estimate the incident particle's energy. 
The short triggering trapezoid with $\tau_\textrm{shape}=5~\upmu$s and $\tau_\textrm{flat}=1~\upmu$s was primarily used for triggering. 
As the short trapezoid only covered the rising part~($<50~\upmu$s) of the waveforms, the height ratio of the two trapezoids $R_\textrm{short/long}$ measured how fast a pulse rises to its maximum. 
Fast-rising pulses have higher values of $R_\textrm{short/long}$.

\begin{figure}[thb]
    \centering
    \includegraphics[width=\columnwidth]{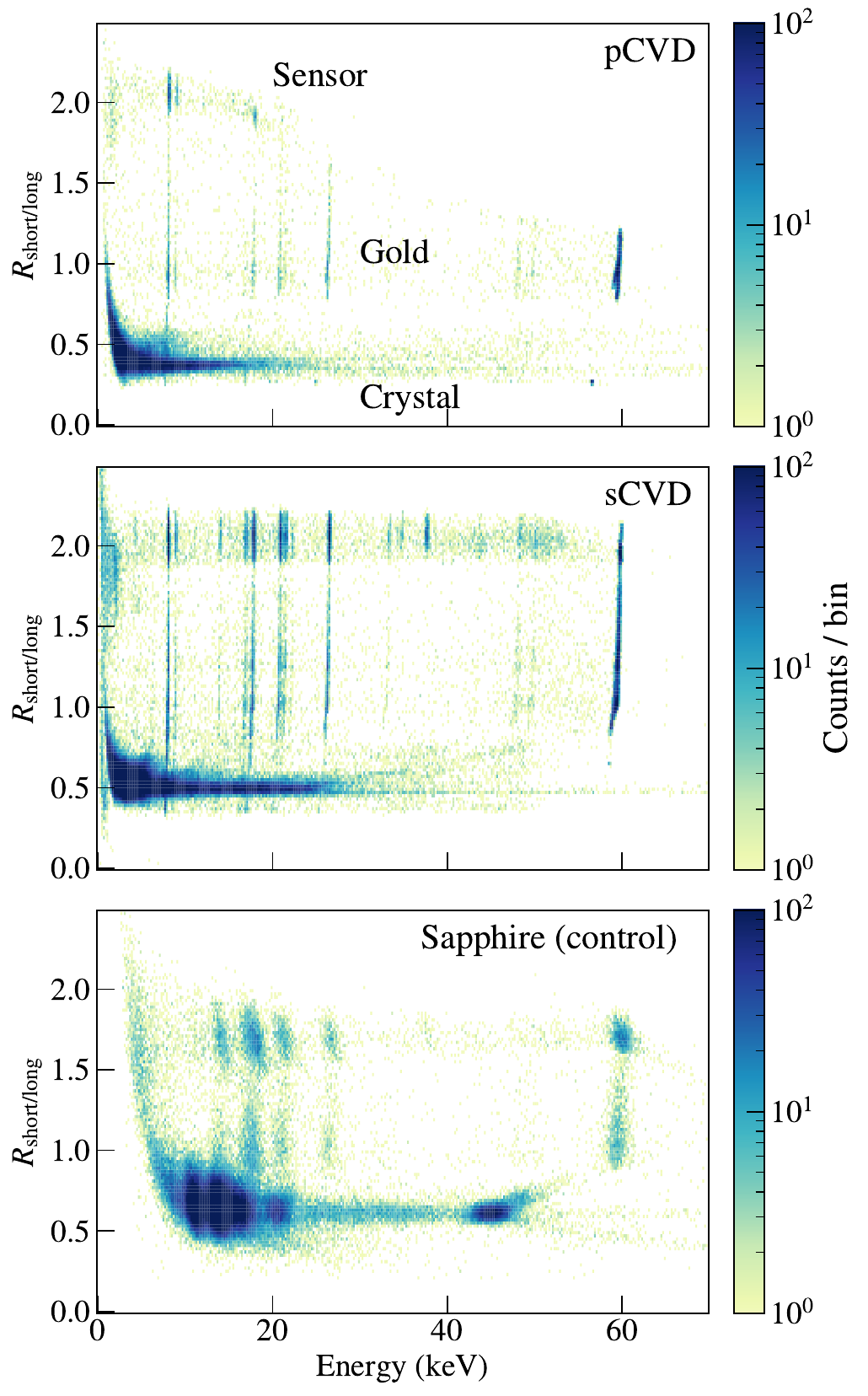}
    \caption{2D histograms showing the ratio of the two trapezoid heights $R_\textrm{short/long}$  vs. energy deposited to the sensor for three different setups, each of which calibrated with their crystal events at $R_\textrm{short/long}\approx1$. (top) result from the pCVD diamond crystal. (middle) result from the sCVD diamond crystal. (bottom) result from the control setup with sapphire absorber.}
    \label{fig:psdScatter}
\end{figure}

\begin{figure}
    \centering
    \includegraphics[width=\columnwidth]{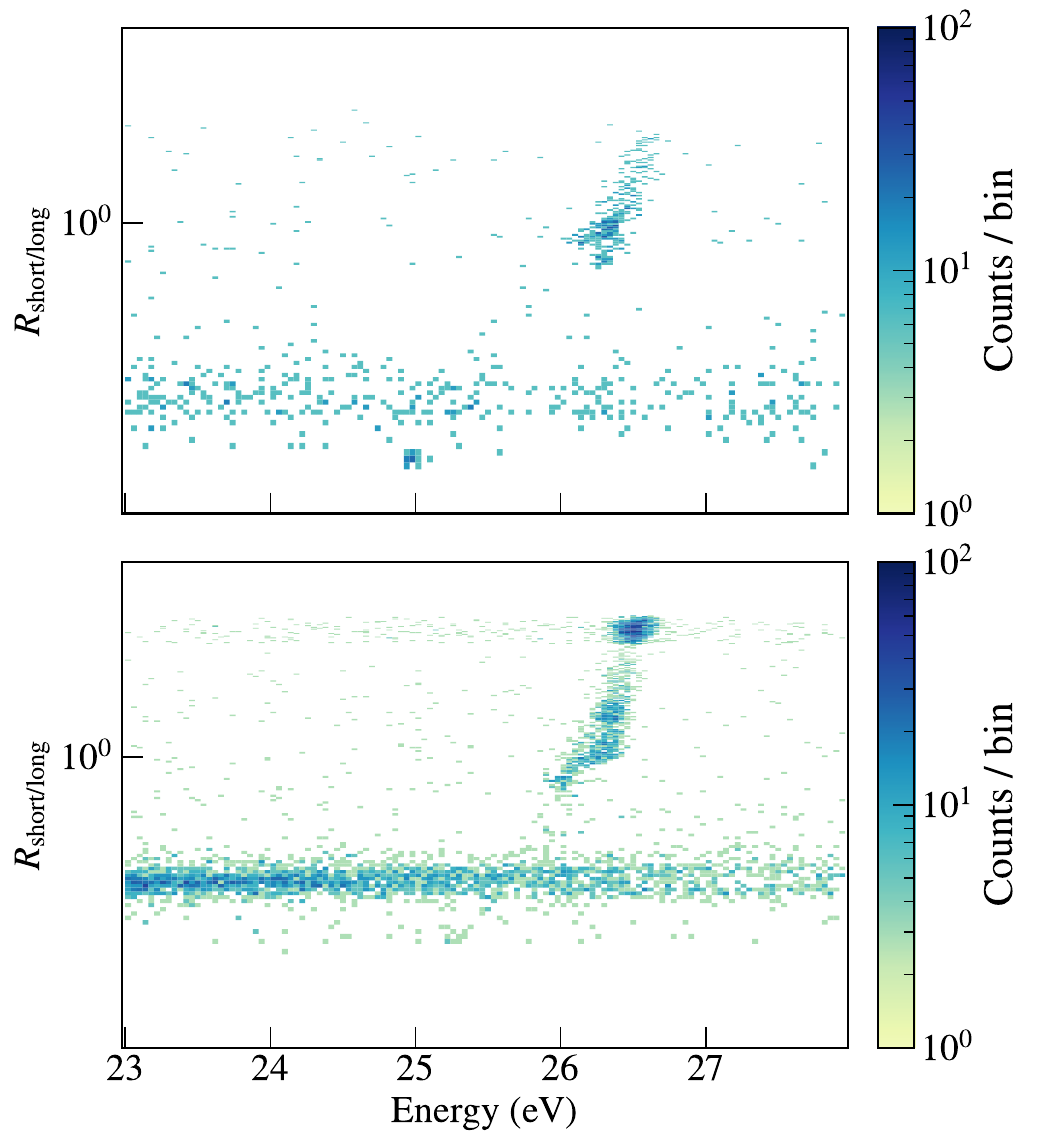}
    \caption{The additional clusters of events observed in the diamond datasets. (top) pCVD data. (bottom)~sCVD data. The new clusters are connected to the gold-hit events, and the pulse amplitudes are $\approx95$\% of the gold-hit events. Counts are normalized by their measurement times.}
    \label{fig:newComponent}
\end{figure}

Figure~\ref{fig:psdScatter} illustrates $R_\textrm{short/long}$  vs. $E$ distribution of events from three different setups: pCVD diamond~(top), sCVD diamond~(middle), and the sapphire control setup (bottom). 
In the sapphire control setup, distinct bands of events at 3 different values of $R_\textrm{short/long}$, at $\approx$~1.8,~1~and~0.8 could be identified. 
Each band exhibits distinct spectral lines from the external $^{241}$Am source as expected, but at different amplitudes.
This is an indication that there existed three main thermal volumes that comprised the detector setup, and that events in the slowest group~(i.e., the group with the lowest values at $R_\textrm{short/long}\approx0.8$) transmit the lowest energy to the sensor. 
We identified each cluster as the sensor hits, the gold phonon absorber hits, and the crystal hits from the fastest to the slowest. 
In the pCVD data, the limitation in the gain-bandwidth-product~(GBP) limited the fastness of the rising component for the larger pulses, and the  sensor hits above $\approx$20~keV were smeared into the gold-hit events band.
In sapphire data, the most prominent 59.54~keV $\gamma-$line in the gold absorber cluster were connected to the crystal-hit cluster at $\approx$45~keV by a diagonal feature, indicating a transition from one volume to another. 
Similar clustering was observed for the diamond crystals as well. 
The events were separated into three distinct bands as in the sapphire setup. 
The diagonal feature similar to what's observed in the sapphire data was again observed in the sCVD data to connect the 59.54~keV gold-hit events and the $\approx$20~keV cluster in the crystal hits, but the feature was not seen in the pCVD data. 
Instead, an additional feature with slower rise times and energy deposits slightly less than the gold hit events was observed in both pCVD and sCVD datasets~(Fig.~\ref{fig:newComponent}). 
These events were again connected to the gold-hit clusters with diagonal components, but at pulse amplitudes of $\approx$95\% of the gold-hit events and rise times even slower than the slow band.

\begin{figure}
    \centering
    \includegraphics[width=\columnwidth]{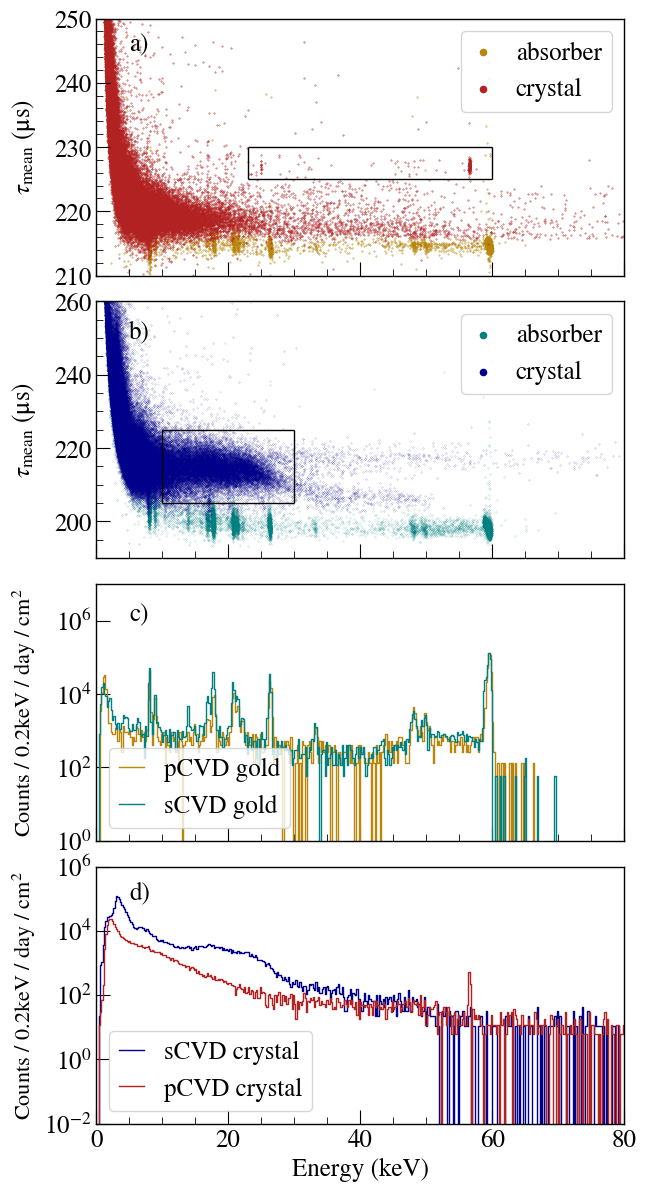}
    \caption{Events distribution in pCVD and sCVD crystals. a) $\tau_\textrm{mean}$ vs energy in pCVD crystal.  b) $\tau_\textrm{mean}$ vs energy in pCVD crystal. Crystal-hit candidates from the $^{241}$Am $\gamma-$sources are selected. c) comparison of gold absorber energy spectra between the pCVD and the sCVD. d) comparison of crystal-hit spectra between the pCVD and the sCVD.}
    \label{fig:spectraComparison}
\end{figure}

The top two plots in Fig.~\ref{fig:spectraComparison} compare the mean-time ($\tau_\textrm{mean}$ ) vs. energy distributions in the two high-resolution diamond datasets. 
The difference in the absolute values of $\tau_\textrm{mean}$ originated from the difference in the filtering conditions. 
The pCVD crystal (Fig.~\ref{fig:spectraComparison} a)) exhibited the two distinct clusters of events at energies slightly lower than 59.54~keV and 26.34~keV, with $\tau_\textrm{mean}\approx228~\upmu$s clearly separated from the band at $\tau_\textrm{mean} \approx 220~\upmu$s. 
The count ratios $A_\textrm{crystal}/A_\textrm{absorber}$ at 59.54~keV and 26.34~keV were 0.045(4) and 0.06(1), respectively. 
The ratio between the two numbers, 0.045(4)/0.06(1)$\approx 0.75\pm0.14$, agreed with the ratio derived from Table~\ref{tab:interactionProbability}, (0.1/0.2)/(1.6/2.2) $\approx 0.7$.
In the sCVD high-resolution dataset, an increase in the event rate at $\approx 20~$keV was identified, at the mean-time value $\tau_\textrm{mean}\approx 215~\upmu$s which was identical to the high-energy band. 
The bottom two plots in Fig.~\ref{fig:spectraComparison} show the energy spectra of the gold absorber hits (c)) and the crystal-hit candidates~(d)), normalized by the exposure and their areas.
The gold absorber hit spectra from the two high-resolution datasets agreed well.
For the crystal hits, the spectra agreed well above 40~keV except at $\approx 58~$keV, where the pCVD data exhibited a peak structure.
In the lower energy region <25~keV, the sCVD data showed an increase. 
We therefore conclude that the majority of the pCVD crystal hits produce slower events with high phonon collection efficiency, while the sCVD crystal hits produce signals with lower phonon collection efficiency and inseparable from the high-energy band. 

\begin{figure}
    \centering
    \includegraphics[width=\columnwidth]{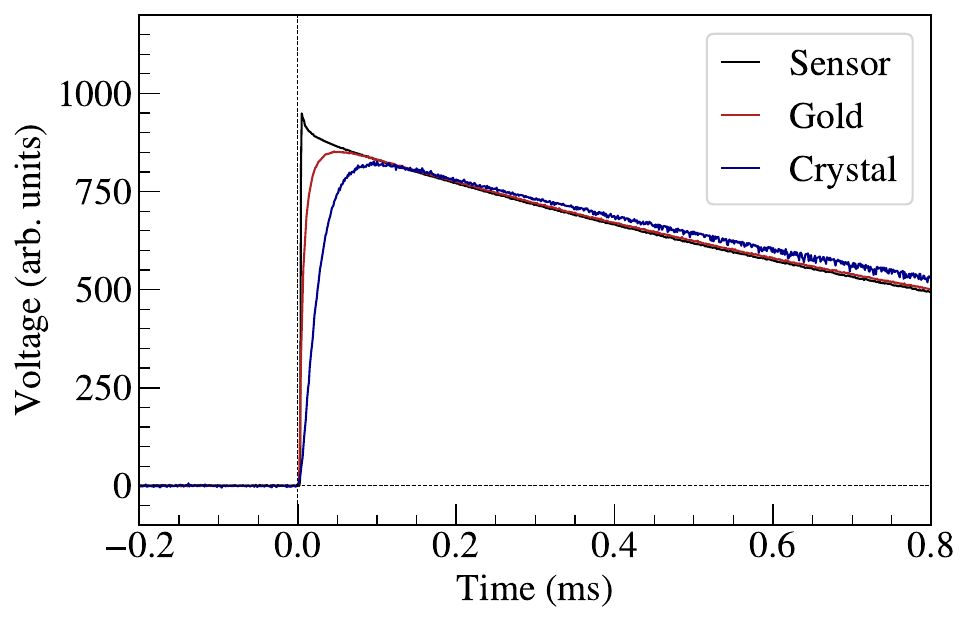}
    \caption{Pulse shapes of the 59.54~keV $\gamma$-ray events hitting different parts of the pCVD detector setup. }
    \label{fig:pulseShapes}
\end{figure}

Similar features appeared in the minimal-filtering datasets as well. Figure~\ref{fig:pulseShapes} features the shapes of pulses from different clusters in the pCVD dataset as a representative. 
We selected the 59.54~keV $\gamma-$lines from the sensor hits, the gold hits, and the crystal hits; and averaged pulses to examine the pulse shape differences. 
The pulses had similar decay times, while having significantly different rising parts. 
This indicates that they follow the same heat flow model as previously explained, and the new features in the diamond crystals also originated from the diamond crystals. 

The fast response of the MMC sensor in combination with the long phonon lifetime inside the diamond crystal would also enable the particle discrimination between $\alpha$ and $\beta/\gamma$ events as explained above, which is not covered in this work.

\subsection{Phonon evolution inside the crystal}

The collection efficiency of phonons created inside the crystal can be estimated to first order by comparing the pulse amplitude ratios of the events depositing energy to the crystal and the events depositing the same amount of energy directly to the absorber. 
This first-order approximation is valid as the decay time in our setup is much longer than the rise time. 
For example, the ratio of the crystal hit amplitudes to the gold hit amplitudes at 59.5~keV, $E_{\mathrm{crystal}}/E_{\mathrm{gold}}(59.5\mathrm{keV})$, was much larger for the pCVD than for the sCVD and for the sapphire. 
Hence we know that the phonon collection efficiency is greater for the pCVD. 
However, the shape differences between the crystal hits and the gold hits cause second-order effects that hinder the accurate calculation of the phonon collection efficiency. 
To fully account for the phonon evolution inside the crystal at finite time scales, the full phonon dynamics in the detector should be considered. 
This can be achieved by deconvolving the pulse shape of the crystal hit events by the detector response.  

\begin{figure}
    \centering
    \includegraphics[width=\columnwidth]{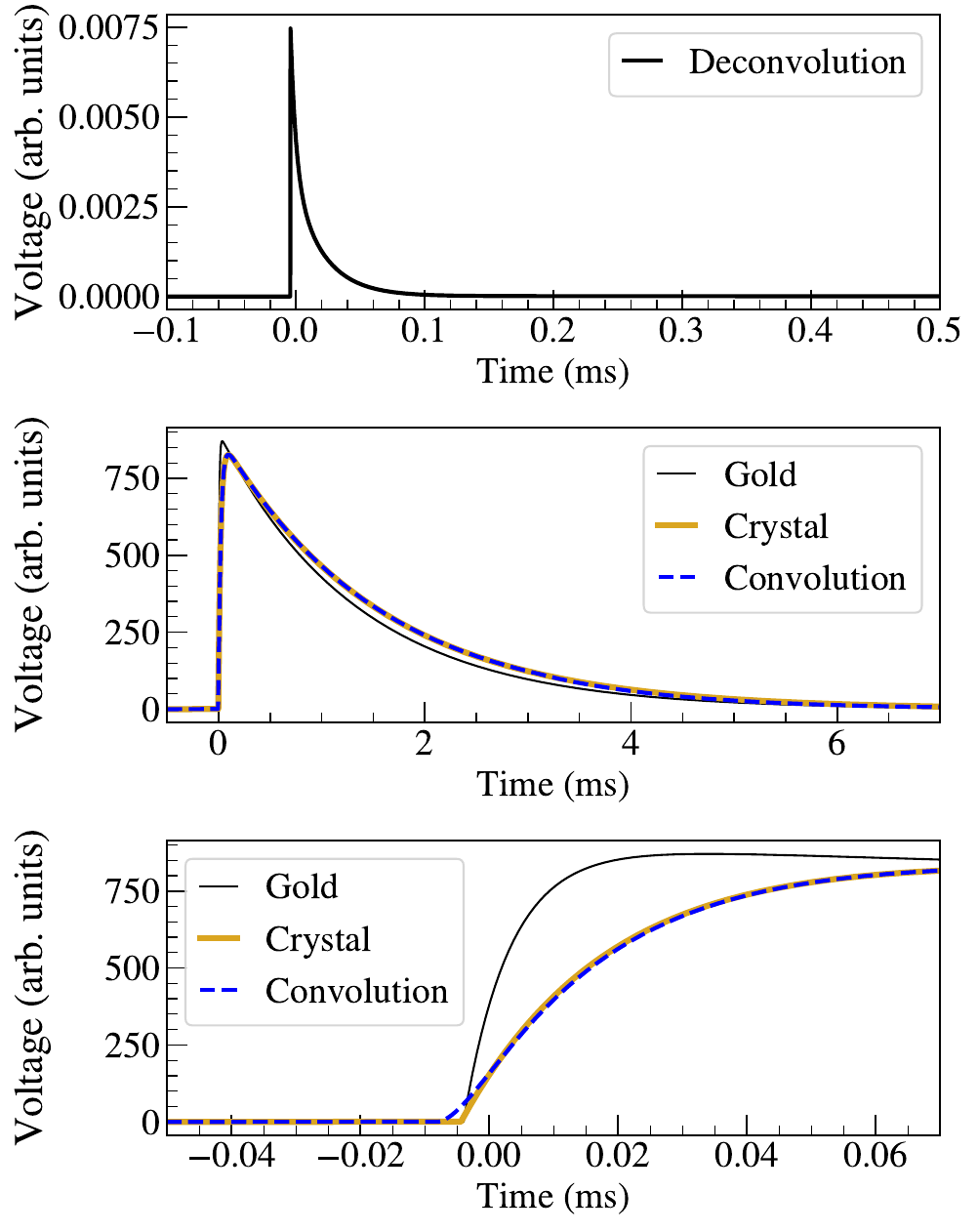}
    \caption{Phonon arrival information extracted by deconvolution from pCVD minimal-filtering data. (top) phonon evolution inside the crystal obtained from deconvolution. (middle) Comparison of the gold-hit and the crystal-hit event, superimposed with the signal reconstruced by convolution. (bottom) the same plot zoomed into the rising edge. The small deviation at pulse onset is due to the limited detector bandwidth.}
    \label{fig:deconvolution}
\end{figure}

In practice, extracting the phonon evolution information with this method is not trivial, as the deconvolution process is strongly affected by the presence of high-frequency noise which persists even after the averaging. 
We therefore conducted a preliminary study by first fitting the waveforms to analytic functions and deconvolving the fit functions to extract the shape information on the phonon evolution. 
Figure~\ref{fig:deconvolution} illustrates an example of this process. 
The deconvolved waveform gives the information about the phonon arrival at the gold phonon absorber. 
The shape parameters extracted from this preliminary analysis were used as initial fit parameters in the final efficiency calculation.

\begin{figure}
    \centering
    \includegraphics[width=\columnwidth]{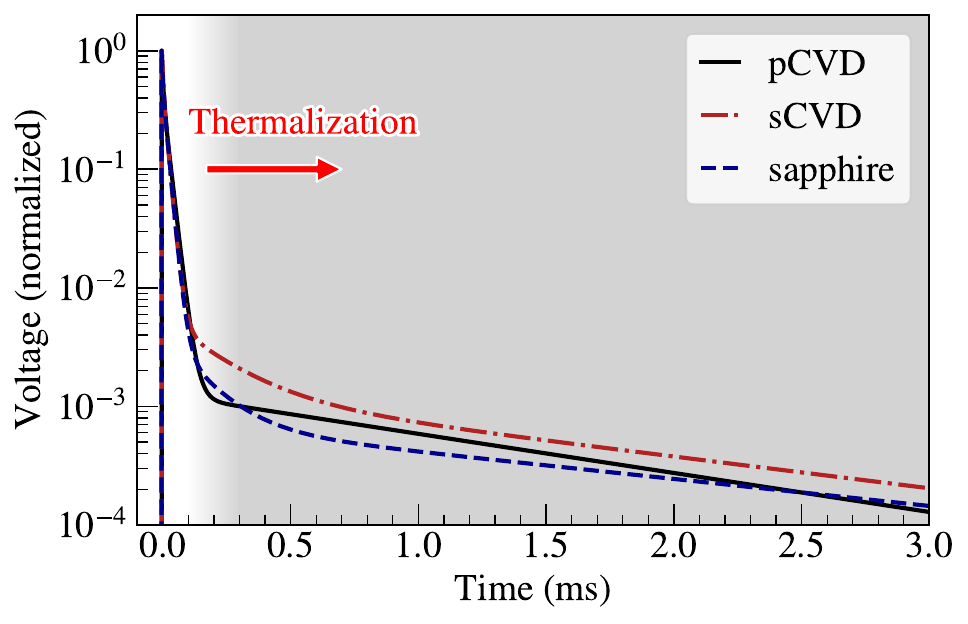}
    \caption{Phonon evolution model inside crystals. There are clearly separable fast and slow components in all three crystals. }
    \label{fig:thermalization}
\end{figure}

In Fig.~\ref{fig:thermalization}, we compared the phonon arrival times in three different crystals. 
Each crystal exhibited a distinct transition from a fast component to a slower component near 200~$\upmu$s. 
The fast component, characterized by a prompt rise, represents the immediate phonon transmission induced by the energy deposition of an incident particle. 
High-energy athermal phonons have high interaction probability with electrons within the phonon absorber, generating a prompt thermal signal that is read by the MMC sensor~\cite{probst1995model}.
On the other hand, the slow component indicates that some phonons are absorbed by the detector with a delay of a few milliseconds, corresponding to thermalized phonons with much weaker interactions with electrons in the metal film~\cite{probst1995model}.
Although the contribution of the thermal component is small compared to the prompt component, it has a significant impact on the waveform when convolved.

In this study, we focused on the analysis of the athermal phonon contribution. 
Absolute quantification of the thermal component is challenging as the contribution of the thermal component in the raw waveform is at a similar frequency to the low-frequency acoustic noise present in our data. 
The thermal phonon analysis may be affected by the 100~Hz high-pass filter that was applied to reduce the vibration-induced low-frequency noise. 
An improved setup, with significantly reduced vibrational noise, a carefully adjusted detector response, and a reduced source rate to minimize pileups, will help accurately quantifying the thermal phonon contribution. 

\begin{figure}
    \centering
    \includegraphics[width=\columnwidth]{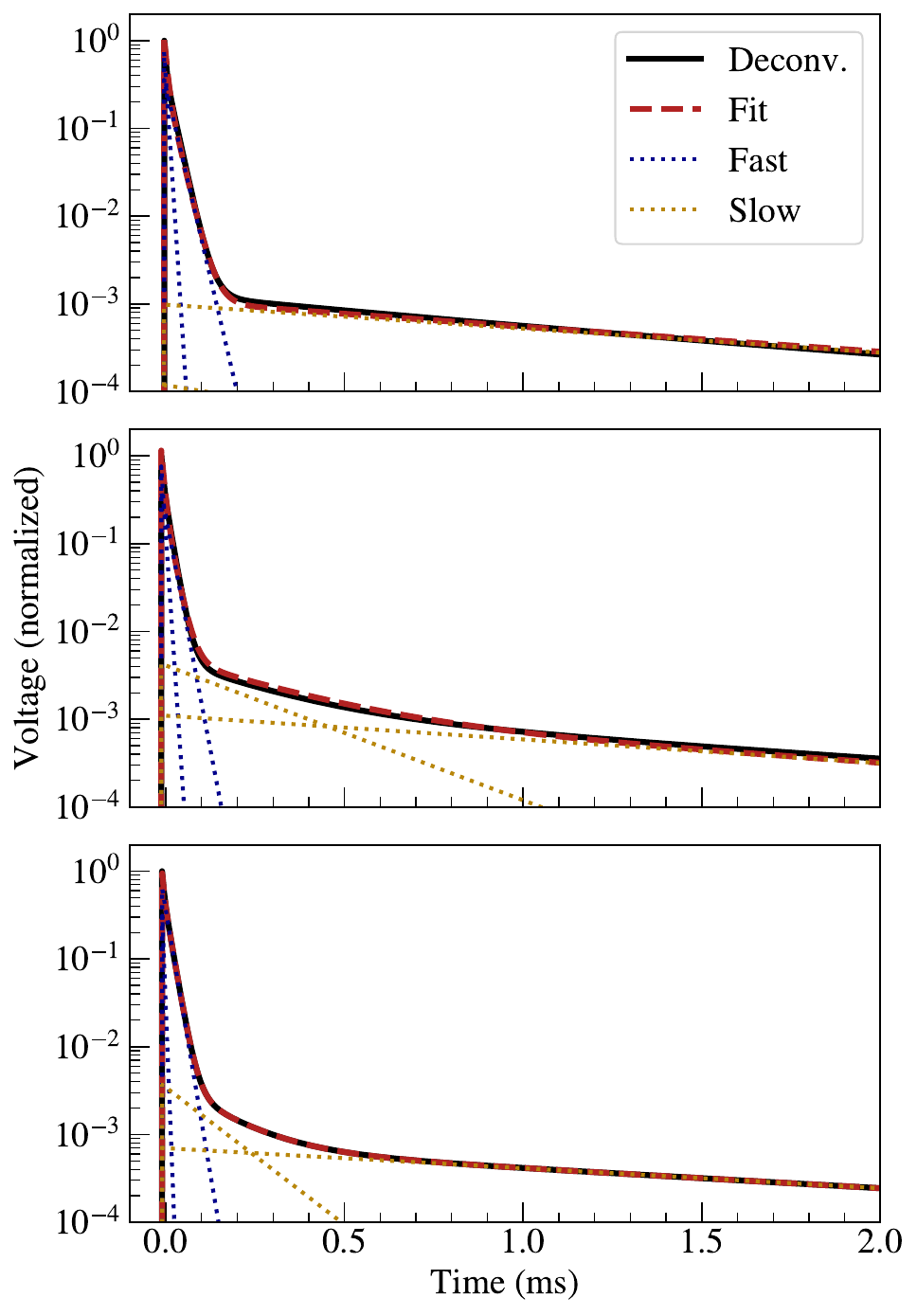}
    \caption{Phonon evolution inside the crystals. (top) the pCVD diamond crystal. (middle) the sCVD diamond crystal. (bottom) the control setup with sapphire absorber. }
    \label{fig:phononDeconvolution}
\end{figure}

The deconvolved phonon evolution can be modelled using a linear combination of four exponentially decaying components: 

\begin{equation}
\begin{split}
    V(t>\tau_0)&=\sum_{i=1}^4 A_i\mathrm{exp}\Big(\frac{(t-\tau_{0})}{\tau_{\mathrm{d},i}}\Big) ~,
\end{split}
\end{equation}

\noindent
where $\tau_{0}$ is the pulse onset time, $\tau_{\mathrm{d},i}$ is the $i-$th decay time constants, and $A_i$ is the amplitude of that component.   
The result of the fit to the model is illustrated in Fig.~\ref{fig:phononDeconvolution}. 
We defined the two fast components with decay times below 100~$\upmu$s as the athermal phonon contribution and the two slow components to be the thermal phonon contribution. 

\begin{figure}
    \centering
    \includegraphics[width=\columnwidth]{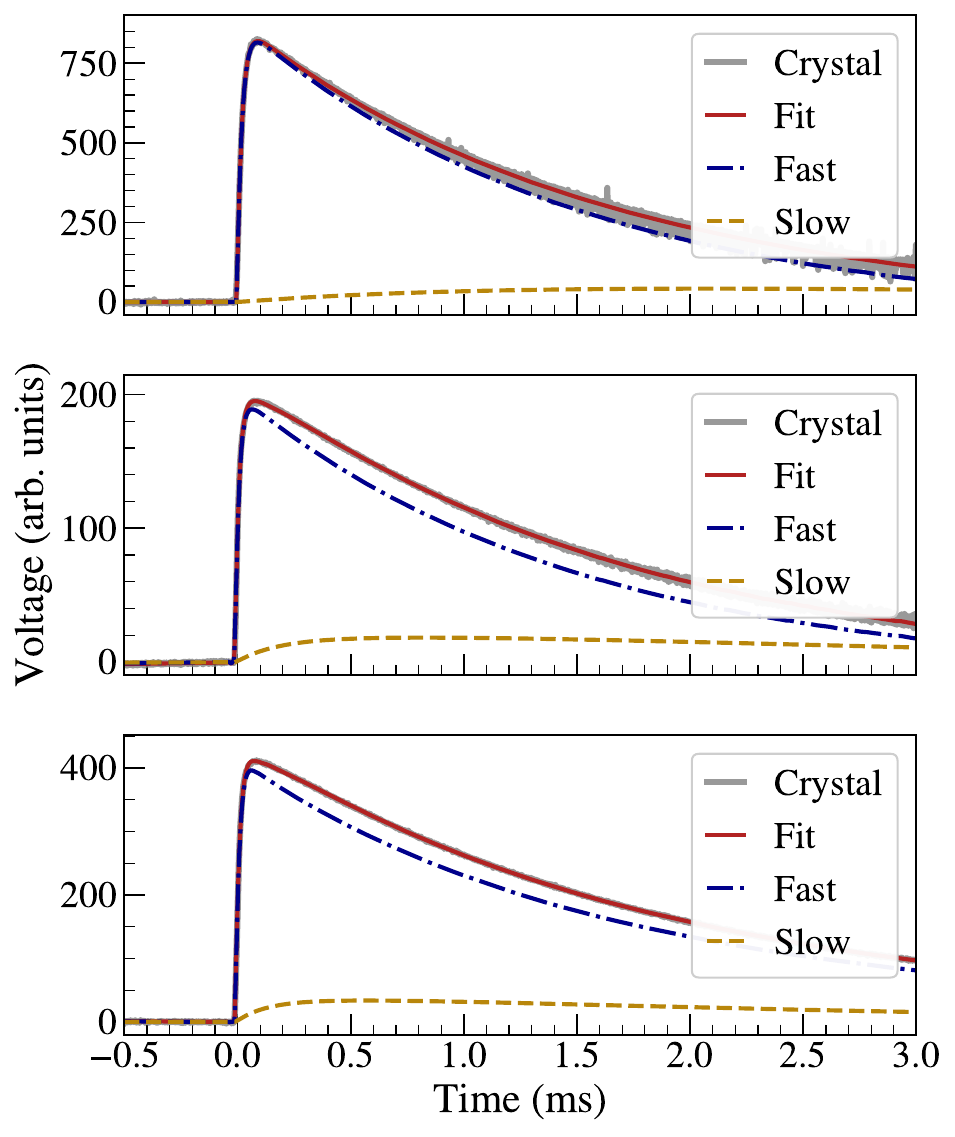}
    \caption{Comparison of the phonon evolution. (top) the pCVD diamond crystal. (middle) the sCVD diamond crystal. (bottom) the control setup with sapphire absorber.}
    \label{fig:convolution}
\end{figure}

Utilizing the results from this preliminary study, we extracted the phonon evolution information from the raw waveforms. 
We found the set of phonon evolution parameters that reproduced the crystal-hit waveforms when convolved with the gold-hit waveform. 
The averaged gold-hit waveform was used as the kernel to minimize the effect of noise, and the convolved waveforms were fit to 20 single crystal-hit waveforms by minimizing $\chi^2$.
Pulses were wavelet-denoised to mitigate the high-frequency noise without affecting the waveforms. 
By convoluting the athermal and the thermal phonon evolutions separately with the detector response again, we modeled the energy flow purely due to the individual components, as illustrated in Fig.~\ref{fig:convolution}.

\begin{table}[t]
    \centering
    \begin{tabular}{|c|c|c||c|}
    \hline
     & pCVD& sCVD &Sapphire\\
     \hline
    Dimension (mm$^3$) & $10\times 10\times 0.5$& $5\times 5\times 0.3$& $5\times 5\times 0.5$ \\
    \hline
    $\mathcal{E}_\mathrm{athermal}$ (\%)& 99.8$\pm$2.3& 39.3$\pm$2.7&73.1$\pm$2.7\\
 $\mathcal{E}_\mathrm{height}$ (\%)& 95.4$\pm$0.5& 45.6$\pm$3.2&76.2$\pm$1.7\\
 $\mathcal{E}_\mathrm{area}$ (\%)& 99.6$\pm$0.6& 45.6$\pm$4.8&83.8$\pm$2.2\\
    \hline
    \end{tabular}
    \caption{Phonon collection efficiency calculated using three different methods.}
    \label{tab:efficiency}
\end{table}

Since the area $A$ of a heat pulse represents the total energy deposit, the athermal phonon collection efficiency $\epsilon_\mathrm{athermal}$ can be calculated by:

\begin{equation}
    \epsilon_\mathrm{athermal}=\frac{A_\mathrm{athermal}}{A_\mathrm{gold}}~,
\end{equation}

\noindent
where $A_\mathrm{athermal}$ is the area of the 'fast' component and $A_\mathrm{gold}$ is the total area of the gold-hit event for equal energy deposited.
The deficit in the efficiency represents the energy loss. 
Table~\ref{tab:efficiency} summarizes the athermal phonon collection efficiencies of different setups. 
The first-order approximations using the pulse amplitudes~$\mathcal{E}_\mathrm{height}$, and using the total area~$\mathcal{E}_\mathrm{area}$,~ are also presented for comparison. 
One can see that the athermal phonon collection efficiency of the pCVD diamond was the largest among the three samples we tested, in spite of the large crystal size.
To the first order, the total phonon collection efficiency calculated using the deconvolution method matched well with the simple approximations. 
In theory, the collection efficiency estimation by the area should account for both the athermal and the thermal contributions in an ideal setup, but this efficiency also approaches towards the athermal phonon collection efficiency in practice, as the contribution from slow thermal phonon is diffused by the limited detector bandwidth, the presence of the noise, and the applied high-pass filter for the acoustic noise mitigation. 

The huge difference in the collection efficiency between the two diamond setups shows that the athermal phonon collection efficiency does not only depend on the chemical properties of the material, but also on the production process, or more generally, on the quality of the crystal. 
Thus, the low temperature detector setup using the MMC sensors could be a direct way of testing the phonon properties and the quality of various crystals with a fast turnaround.
Our results re-verifies the previously reported results on the diamond detector studies, that the pCVD can exhibit good properties comparable to the sCVD crystals if the crystal quality is high~\cite{liu2017diamond,gallin2021characterization} .
This could be an evidence that the phonon transmission is affected more by the crystal quality than by the grain boundaries. 

\subsection{Low-energy measurement}

The high phonon collection efficiency observed from the pCVD diamond, which led to a high energy resolution, enables a diamond-based detector to detect recoil signals at $\mathcal{O}(10)$~eV energy range.
In combination with the low nuclear mass of the carbon atoms forming the crystal, the diamond-based low temperature detector can be sensitive to nuclear recoils induced by sub-GeV dark matter particles down to $\mathcal{O}(100)$~MeV.

We examined the energy resolution of the diamond-based detector using the 8.05 keV copper fluorescence peak in the high-resolution dataset. 
Since the MMC sensors measure heat flow without being affected by quenching, the NR and the ER signals in our measurements follow a similar calibration curve. 
While there can be slight deviations in the energy scale between ER and NR signals if the electron-hole pair lifetime significantly exceeds the detector response speed, this factor is considered to be of second-order and was not considered in this work.

To achieve the best energy resolution using the current setup, we processed the pCVD high-resolution dataset with an optimum filter, using the averaged 59.5~keV $\gamma$ events absorbed in the crystal as the template. 
The filtered signal had the maximum signal-to-noise ratio, which improved both the triggering threshold and the energy resolution. 

To reduce the amount of noise-induced false trigger, we implemented a high-frequency rejection algorithm on the triggered waveform. 
Once the waveform was triggered with the optimum filter, we examined five sampling points in the flat-top region of the long trapezoid waveforms~(10\%, 30\%, 50\%, 80\% and 90\% points) and only accepted waveforms if all five points exceeds the threshold. 
In addition, if the mean time of the raw waveform appeared outside of the expected center of mass region of the waveform, we considered the waveform to be electronics noise and veto them. 
These high-frequency algorithms suppressed the false-trigger signals at the cost of reduced detection efficiency. 
The final energy was calculated after the correction for the temperature drifts and the pulse shape dependence of the energy estimator. 

\begin{figure}
    \centering
    \includegraphics[width=\columnwidth]{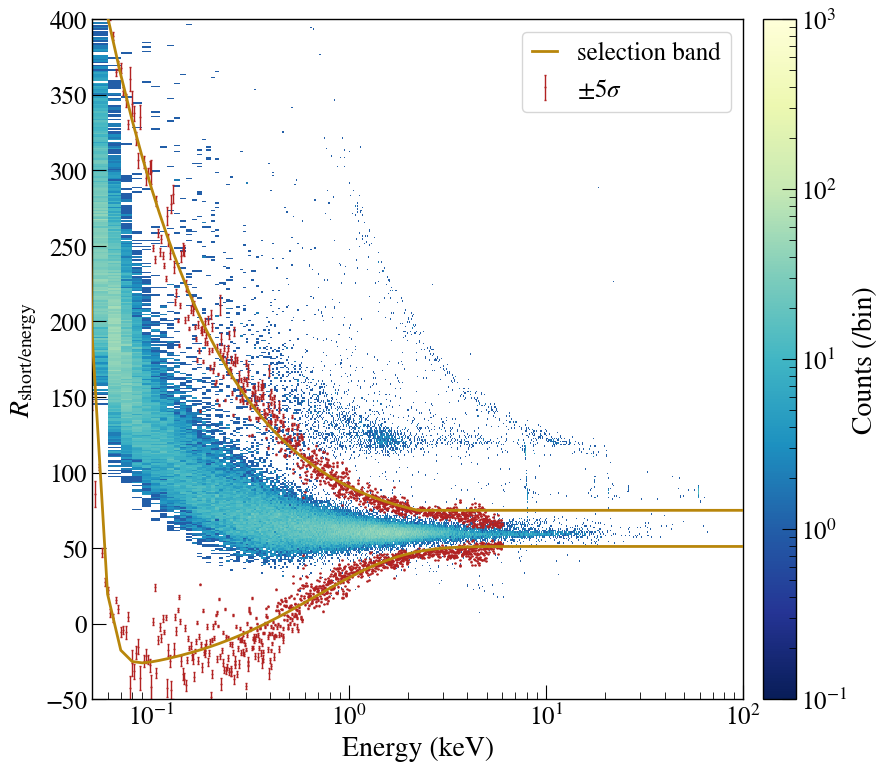}
    \caption{
    Fastness parameter $R_\textrm{short/energy}$ vs. Energy scatter plot and the $\pm5\sigma$ selection band for the crystal-hit events.
    This plot is similar to Fig.~\ref{fig:psdScatter}, but both parameters are calculated from the optimal-filter derived amplitude instead of the long trapezoid for better resolution.
    }
    \label{fig:fastness}
\end{figure}

Figure~\ref{fig:fastness} shows the fastness parameter $R_\textrm{short/energy}$ versus Energy scatter plot after correction. 
$R_\mathrm{short/energy}$ was analogous to $R_\mathrm{short/long}$ used previously, but used the optimal-filter derived amplitude in the determination of the ratio.
To define the $\pm5\sigma$ selection band for the crystal-hit events, the values of $R_\textrm{short/energy}$ at various energy slices were fit to Gaussian distributions using Random Sample Consensus~(RANSAC) algorithm~\cite{ransac1981} to minimize distortions from other bodies. 
We examined the four most prominent x-ray peaks from $^{241}$Am source at 17.8~keV, 26.34~keV and 59.54~keV, and the 8.05~keV copper fluorescence peak, and applied a quadratic calibration with zero constant to calibrate the spectrum.

\begin{figure}
    \centering
    \includegraphics[width=\columnwidth]{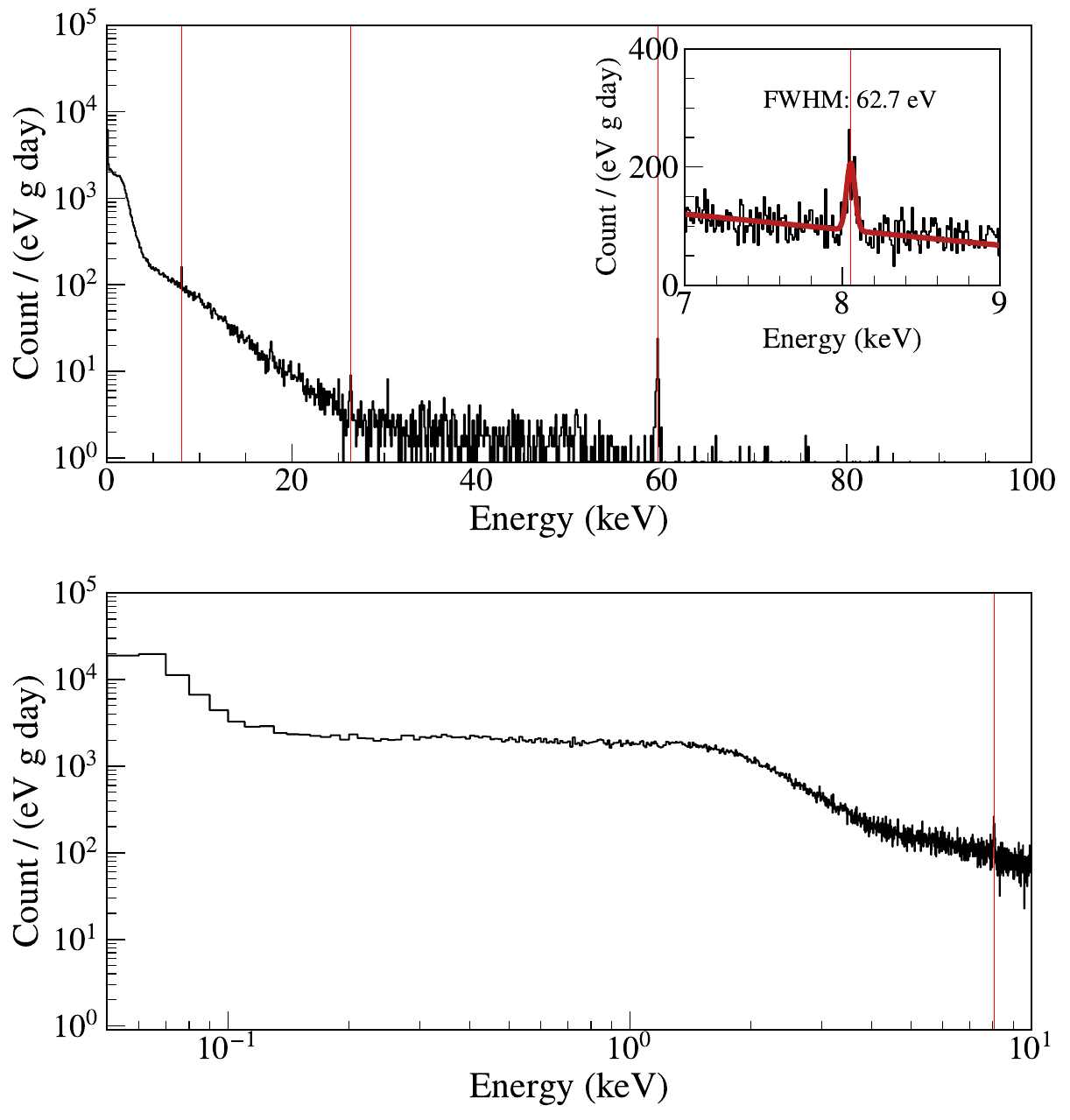}
    \caption{Final energy spectra from crystal-hit events. (top) up to 100~keV. Red vertical lines indicate 8.05~keV, 26.34~keV and 59.54~keV photopeaks. The inset illustrates the 8.05~keV photopeak with a Gaussian fit. The FWHM at 8.05~keV is 62.7~eV (0.78\%). (bottom) low-energy spectrum up to 10~keV in logarithmic x-axis.}
    \label{fig:finalSpectrum}
\end{figure}

\begin{figure}
    \centering
    \includegraphics[width=\columnwidth]{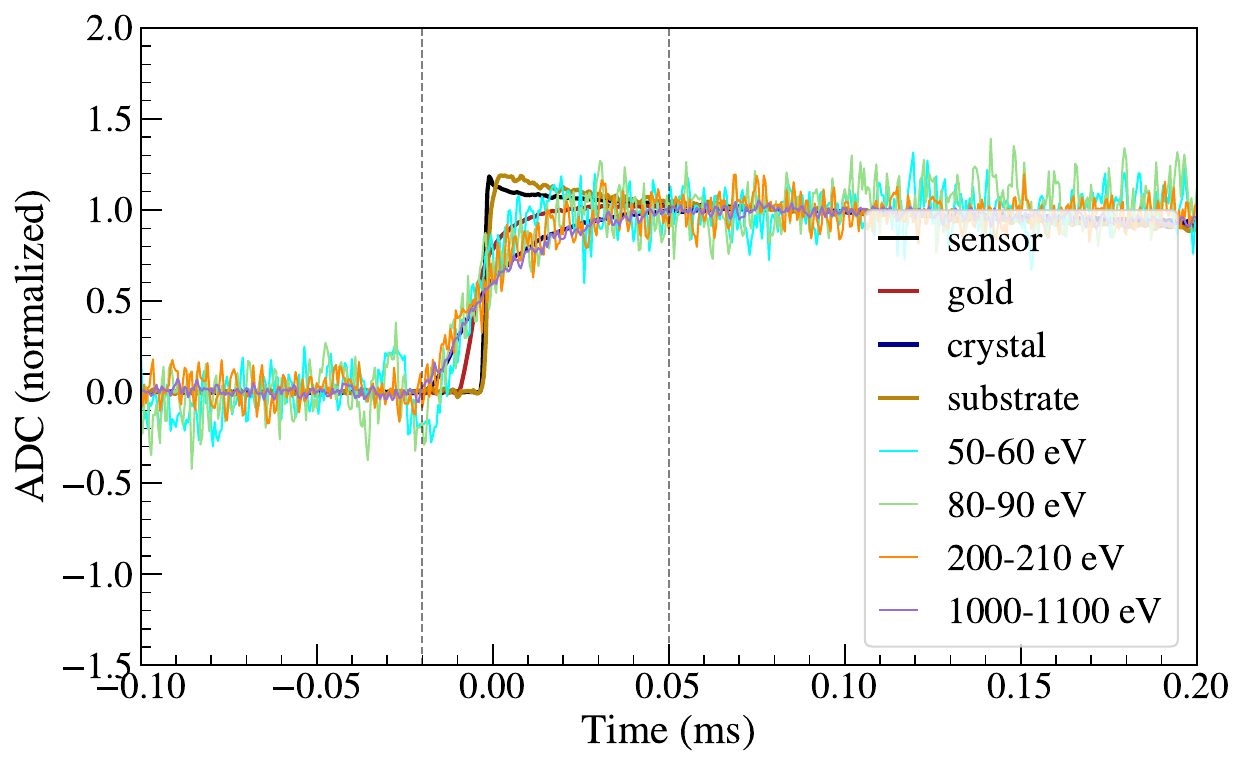}
    \caption{
    Average waveforms at 4 different energies, normalized by their energies. 
    Higher-energy waveforms from sensor hits, gold absorber hits, crystal hits~(8~keV), and substrate hits~(4~keV) are overlaid for comparison.
    The low-energy waveforms are closer in shape to those of gold absorber or crystal hit events than those of substrate or sensor hits.
    Gray dashed lines indicate the rising part of the waveforms. 
    }
    \label{fig:avgWaveforms}
\end{figure}

Figure~\ref{fig:finalSpectrum} shows the final energy spectra from the crystal hit events. 
We found the FWHM energy resolution at 8.05~keV to be 62.7~eV~(0.78\%). 
An increase in number of events below 150~eV was observed in the low-energy spectrum. 
To verify these are not noise-induced false-triggers, we averaged waveforms at different energies and compare their shapes in Fig.~\ref{fig:avgWaveforms}.
Only waveforms with flat baselines were included in the averaging.
Higher-energy waveforms corresponding to sensor hits, gold absorber hits, crystal hits~(8~keV) and substrate hits~(4~keV) are overlaid for comparison.
The comparison of the waveforms normalized by their energies clearly demonstrates that they do have similar rise-times, indicating they originate from the similar heat-flow processes.

\begin{figure}
    \centering
    \includegraphics[width=\columnwidth]{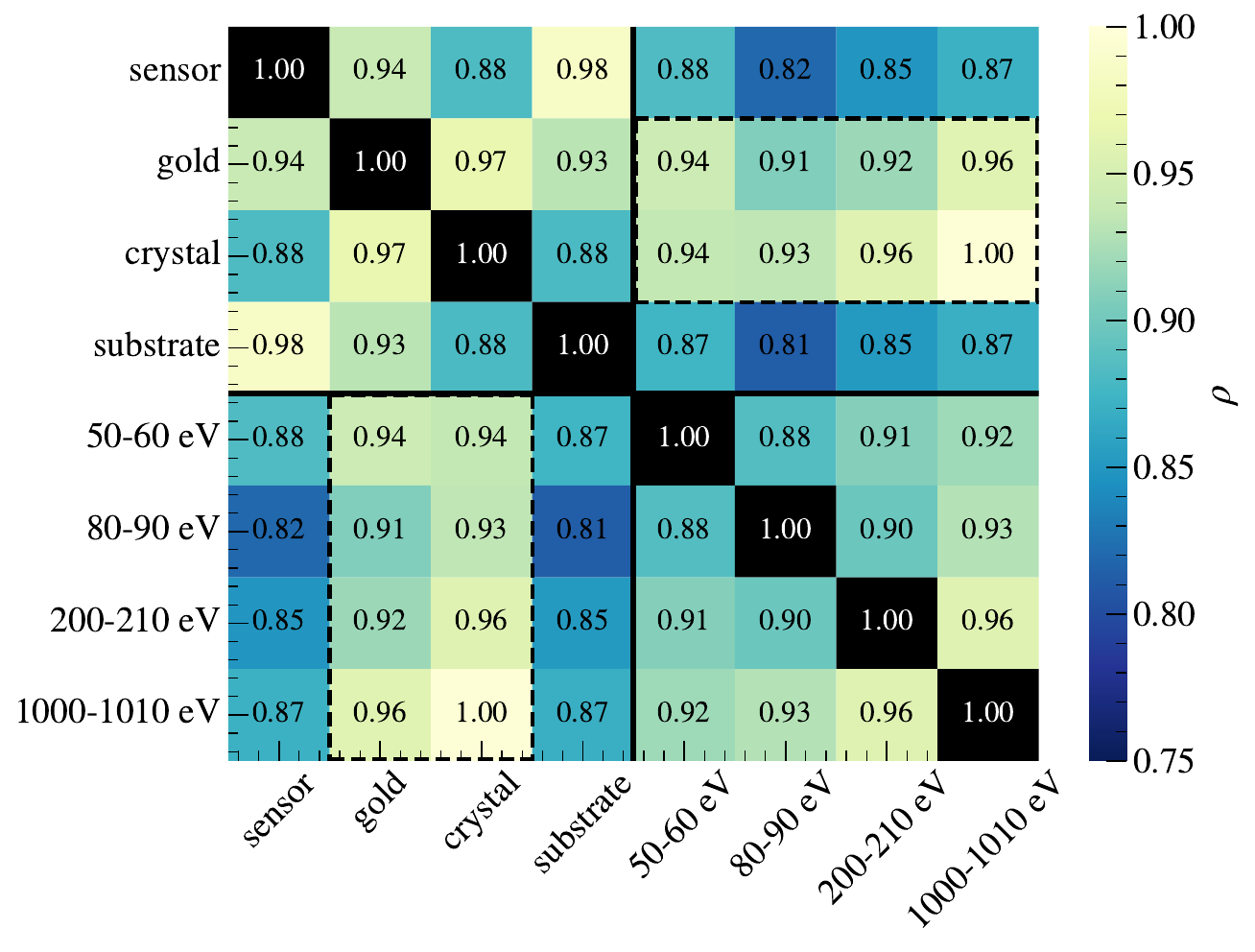}
    \caption{
    Pearson's correlation coefficient, $\rho$, calculated for the rising part of the waveforms. 
    The gray dashed lines at -0.02~ms and 0.05~ms indicate the time range used to calculate $\rho$.
    The lower-energy waveforms are closer in shape to gold or crystal hits, compared to  sensor or substrate hits.
    }
    \label{fig:CorrelationMtx}
\end{figure}

The pulse shape comparison is quantified using Pearson's correlation coefficient, $\rho$, as shown in Fig.~\ref{fig:CorrelationMtx}. 
The analysis was done on the rising part of the waveforms,  in the time interval -20$\upmu$s < t < 50$\upmu$s~(Fig.~\ref{fig:avgWaveforms}). 
Dashed boxes indicate regions of strong correlation between the low-energy waveforms and the higher-energy templates. 
In every case, the lower energy waveforms show higher correlation coefficients with either gold absorber hits or crystal hits than with sensor or substrate hits. 
This indicates that the low-energy events are from either gold hits or crystal hits rather than from substrate or sensor hits.
However, with the current experimental precision, distinguishing between gold absorber hits and crystal hits is challenging.
Future measurements using the fast phonon sensors of MAGNETO-DM, with sub-microsecond response times, may enable more detailed study on the origin of these events.

The undershoots prior to the low-energy waveforms indicates the presence of high-frequency noise.
As the low-energy waveforms were comparable in amplitudes with the high-frequency noise, they could only surpass the trigger level if they were in-phase with the high-frequency noise. 
This phase selection was enhanced when averaging the waveforms and appeared as the undershoot just before the rising edge. 
This suggests that the trigger efficiency was reduced due to the presence of the noise at low-energy region. 

These results clearly demonstrate the existence of the increase in counts at low energy region below 100~eV, typically known to the community as the low-energy excess~(LEE)~\cite{excess2022}. 
This feature would become more prominent if the low detection efficiency at the low energy region is considered. 
If the detector is upgraded to the direct coupling design as explained above, the MMC-based detector will have enough bandwidth to both measure recoil with even lower energies and extract the thermal and athermal components of the events comprising the LEE. 
This will provide valuable information about the nature of the LEE, telling us whether the LEE is of thermal origin or non-thermal origin. 

\section{Conclusion}
We conducted tests to assess high-quality lab-grown diamond crystals as particle absorbers for future dark matter searches using the MMC phonon sensors. 
We developed a novel method of characterizing the phonon evolution inside the target crystal by exploiting the high resolution and fast response of the MMC magnetic quantum sensors. 
The crystal fabricated via the polycrystalline chemical vapor deposition~(pCVD) technique appeared to meet the stringent quality criteria for the DM search experiments. 
The phonon collection efficiency was found to be as high as $(99.8\pm2.3)$\% in the best setup.

With the pCVD diamond sample, we achieved energy resolution of 61.9~eV at 8.05~keV. 
The energy resolution is expected to improve further using the advanced direct-contact MMC design that is currently under development. 
The analysis of the high-resolution dataset, however, also illustrated that some spectral features in the data are not yet fully understood. 
In particular, the origin of the band of events at $\tau_\textrm{mean}\approx218~\upmu$s in the pCVD data~(Fig.~\ref{fig:spectraComparison}) is not yet known. 
Future measurements with the fast phonon sensors of MAGNETO-DM with sub-microsecond response times may allow to investigate their origin.

With the energy threshold lower than 100~eV, we confirmed in our setup the existence of the low-energy excess reported from multiple low-temperature experiments. 
Future study with the diamond crystals and the MMC sensors will provide critical information and pin down the origin of the LEE.

\section{Acknowledgments}
This work was performed under the auspices of the U.S. Department of Energy by Lawrence Livermore National Laboratory under Contract DE-AC52-07NA27344. This work was supported by the Laboratory Directed Research and Development program of Lawrence Livermore National Laboratory (22-FS-011) and DOE Office of Science HEP Advanced Detector R\&D program.

\newpage

\bibliographystyle{apsrev4-1}
\bibliography{00ref}

\begin{thebibliography}{70}%
\makeatletter
\providecommand \@ifxundefined [1]{%
 \@ifx{#1\undefined}
}%
\providecommand \@ifnum [1]{%
 \ifnum #1\expandafter \@firstoftwo
 \else \expandafter \@secondoftwo
 \fi
}%
\providecommand \@ifx [1]{%
 \ifx #1\expandafter \@firstoftwo
 \else \expandafter \@secondoftwo
 \fi
}%
\providecommand \natexlab [1]{#1}%
\providecommand \enquote  [1]{``#1''}%
\providecommand \bibnamefont  [1]{#1}%
\providecommand \bibfnamefont [1]{#1}%
\providecommand \citenamefont [1]{#1}%
\providecommand \href@noop [0]{\@secondoftwo}%
\providecommand \href [0]{\begingroup \@sanitize@url \@href}%
\providecommand \@href[1]{\@@startlink{#1}\@@href}%
\providecommand \@@href[1]{\endgroup#1\@@endlink}%
\providecommand \@sanitize@url [0]{\catcode `\\12\catcode `\$12\catcode `\&12\catcode `\#12\catcode `\^12\catcode `\_12\catcode `\%12\relax}%
\providecommand \@@startlink[1]{}%
\providecommand \@@endlink[0]{}%
\providecommand \url  [0]{\begingroup\@sanitize@url \@url }%
\providecommand \@url [1]{\endgroup\@href {#1}{\urlprefix }}%
\providecommand \urlprefix  [0]{URL }%
\providecommand \Eprint [0]{\href }%
\providecommand \doibase [0]{http://dx.doi.org/}%
\providecommand \selectlanguage [0]{\@gobble}%
\providecommand \bibinfo  [0]{\@secondoftwo}%
\providecommand \bibfield  [0]{\@secondoftwo}%
\providecommand \translation [1]{[#1]}%
\providecommand \BibitemOpen [0]{}%
\providecommand \bibitemStop [0]{}%
\providecommand \bibitemNoStop [0]{.\EOS\space}%
\providecommand \EOS [0]{\spacefactor3000\relax}%
\providecommand \BibitemShut  [1]{\csname bibitem#1\endcsname}%
\let\auto@bib@innerbib\@empty
%</preamble>
\bibitem [{\citenamefont {Clowe}\ \emph {et~al.}(2006)\citenamefont {Clowe}, \citenamefont {Brada{\v{c}}}, \citenamefont {Gonzalez}, \citenamefont {Markevitch}, \citenamefont {Randall}, \citenamefont {Jones},\ and\ \citenamefont {Zaritsky}}]{Clowe_2006_DM}%
  \BibitemOpen
  \bibfield  {author} {\bibinfo {author} {\bibfnamefont {D.}~\bibnamefont {Clowe}}, \bibinfo {author} {\bibfnamefont {M.}~\bibnamefont {Brada{\v{c}}}}, \bibinfo {author} {\bibfnamefont {A.~H.}\ \bibnamefont {Gonzalez}}, \bibinfo {author} {\bibfnamefont {M.}~\bibnamefont {Markevitch}}, \bibinfo {author} {\bibfnamefont {S.~W.}\ \bibnamefont {Randall}}, \bibinfo {author} {\bibfnamefont {C.}~\bibnamefont {Jones}}, \ and\ \bibinfo {author} {\bibfnamefont {D.}~\bibnamefont {Zaritsky}},\ }\href {\doibase 10.1086/508162} {\bibfield  {journal} {\bibinfo  {journal} {The Astrophysical Journal}\ }\textbf {\bibinfo {volume} {648}},\ \bibinfo {pages} {L109} (\bibinfo {year} {2006})}\BibitemShut {NoStop}%
\bibitem [{\citenamefont {Liu}\ \emph {et~al.}(2017{\natexlab{a}})\citenamefont {Liu}, \citenamefont {Chen},\ and\ \citenamefont {Ji}}]{liu2017current}%
  \BibitemOpen
  \bibfield  {author} {\bibinfo {author} {\bibfnamefont {J.}~\bibnamefont {Liu}}, \bibinfo {author} {\bibfnamefont {X.}~\bibnamefont {Chen}}, \ and\ \bibinfo {author} {\bibfnamefont {X.}~\bibnamefont {Ji}},\ }\href {\doibase 10.1038/nphys4039} {\bibfield  {journal} {\bibinfo  {journal} {Nature Physics}\ }\textbf {\bibinfo {volume} {13}},\ \bibinfo {pages} {212} (\bibinfo {year} {2017}{\natexlab{a}})}\BibitemShut {NoStop}%
\bibitem [{\citenamefont {Undagoitia}\ and\ \citenamefont {Rauch}(2015)}]{Undagoitia_2015}%
  \BibitemOpen
  \bibfield  {author} {\bibinfo {author} {\bibfnamefont {T.~M.}\ \bibnamefont {Undagoitia}}\ and\ \bibinfo {author} {\bibfnamefont {L.}~\bibnamefont {Rauch}},\ }\href {\doibase 10.1088/0954-3899/43/1/013001} {\bibfield  {journal} {\bibinfo  {journal} {Journal of Physics G: Nuclear and Particle Physics}\ }\textbf {\bibinfo {volume} {43}},\ \bibinfo {pages} {013001} (\bibinfo {year} {2015})}\BibitemShut {NoStop}%
\bibitem [{\citenamefont {Choi}\ \emph {et~al.}(2015)\citenamefont {Choi}, \citenamefont {Abe}, \citenamefont {Haga}, \citenamefont {Hayato}, \citenamefont {Iyogi}, \citenamefont {Kameda}, \citenamefont {Kishimoto}, \citenamefont {Miura}, \citenamefont {Moriyama}, \citenamefont {Nakahata} \emph {et~al.}}]{choi2015skk}%
  \BibitemOpen
  \bibfield  {author} {\bibinfo {author} {\bibfnamefont {K.}~\bibnamefont {Choi}}, \bibinfo {author} {\bibfnamefont {K.}~\bibnamefont {Abe}}, \bibinfo {author} {\bibfnamefont {Y.}~\bibnamefont {Haga}}, \bibinfo {author} {\bibfnamefont {Y.}~\bibnamefont {Hayato}}, \bibinfo {author} {\bibfnamefont {K.}~\bibnamefont {Iyogi}}, \bibinfo {author} {\bibfnamefont {J.}~\bibnamefont {Kameda}}, \bibinfo {author} {\bibfnamefont {Y.}~\bibnamefont {Kishimoto}}, \bibinfo {author} {\bibfnamefont {M.}~\bibnamefont {Miura}}, \bibinfo {author} {\bibfnamefont {S.}~\bibnamefont {Moriyama}}, \bibinfo {author} {\bibfnamefont {M.}~\bibnamefont {Nakahata}},  \emph {et~al.} (\bibinfo {collaboration} {Super-Kamiokande Collaboration}),\ }\href {\doibase 10.1103/PhysRevLett.114.141301} {\bibfield  {journal} {\bibinfo  {journal} {Phys. Rev. Lett.}\ }\textbf {\bibinfo {volume} {114}},\ \bibinfo {pages} {141301} (\bibinfo {year} {2015})}\BibitemShut {NoStop}%
\bibitem [{\citenamefont {Agnese}\ \emph {et~al.}(2016)\citenamefont {Agnese}, \citenamefont {Anderson}, \citenamefont {Aramaki}, \citenamefont {Asai}, \citenamefont {Baker}, \citenamefont {Balakishiyeva}, \citenamefont {Barker}, \citenamefont {Basu~Thakur}, \citenamefont {Bauer}, \citenamefont {Billard} \emph {et~al.}}]{agnese2016cdms}%
  \BibitemOpen
  \bibfield  {author} {\bibinfo {author} {\bibfnamefont {R.}~\bibnamefont {Agnese}}, \bibinfo {author} {\bibfnamefont {A.~J.}\ \bibnamefont {Anderson}}, \bibinfo {author} {\bibfnamefont {T.}~\bibnamefont {Aramaki}}, \bibinfo {author} {\bibfnamefont {M.}~\bibnamefont {Asai}}, \bibinfo {author} {\bibfnamefont {W.}~\bibnamefont {Baker}}, \bibinfo {author} {\bibfnamefont {D.}~\bibnamefont {Balakishiyeva}}, \bibinfo {author} {\bibfnamefont {D.}~\bibnamefont {Barker}}, \bibinfo {author} {\bibfnamefont {R.}~\bibnamefont {Basu~Thakur}}, \bibinfo {author} {\bibfnamefont {D.~A.}\ \bibnamefont {Bauer}}, \bibinfo {author} {\bibfnamefont {J.}~\bibnamefont {Billard}},  \emph {et~al.} (\bibinfo {collaboration} {SuperCDMS Collaboration}),\ }\href {\doibase 10.1103/PhysRevLett.116.071301} {\bibfield  {journal} {\bibinfo  {journal} {Phys. Rev. Lett.}\ }\textbf {\bibinfo {volume} {116}},\ \bibinfo {pages} {071301} (\bibinfo {year} {2016})}\BibitemShut {NoStop}%
\bibitem [{\citenamefont {Armengaud}\ \emph {et~al.}(2016)\citenamefont {Armengaud}, \citenamefont {Arnaud}, \citenamefont {Augier}, \citenamefont {Beno{\^\i}t}, \citenamefont {Berg{\'e}}, \citenamefont {Bergmann}, \citenamefont {Billard}, \citenamefont {Bl{\"u}mer}, \citenamefont {De~Boissi{\`e}re}, \citenamefont {Bres} \emph {et~al.}}]{armengaud2016edelweiss}%
  \BibitemOpen
  \bibfield  {author} {\bibinfo {author} {\bibfnamefont {E.}~\bibnamefont {Armengaud}}, \bibinfo {author} {\bibfnamefont {Q.}~\bibnamefont {Arnaud}}, \bibinfo {author} {\bibfnamefont {C.}~\bibnamefont {Augier}}, \bibinfo {author} {\bibfnamefont {A.}~\bibnamefont {Beno{\^\i}t}}, \bibinfo {author} {\bibfnamefont {L.}~\bibnamefont {Berg{\'e}}}, \bibinfo {author} {\bibfnamefont {T.}~\bibnamefont {Bergmann}}, \bibinfo {author} {\bibfnamefont {J.}~\bibnamefont {Billard}}, \bibinfo {author} {\bibfnamefont {J.}~\bibnamefont {Bl{\"u}mer}}, \bibinfo {author} {\bibfnamefont {T.}~\bibnamefont {De~Boissi{\`e}re}}, \bibinfo {author} {\bibfnamefont {G.}~\bibnamefont {Bres}},  \emph {et~al.},\ }\href {\doibase 10.1088/1475-7516/2016/05/019} {\bibfield  {journal} {\bibinfo  {journal} {Journal of Cosmology and Astroparticle Physics}\ }\textbf {\bibinfo {volume} {2016}},\ \bibinfo {pages} {019} (\bibinfo {year} {2016})}\BibitemShut {NoStop}%
\bibitem [{\citenamefont {Arnaud}\ \emph {et~al.}(2020)\citenamefont {Arnaud}, \citenamefont {Armengaud}, \citenamefont {Augier}, \citenamefont {Beno\^{\i}t}, \citenamefont {Berg\'e}, \citenamefont {Billard}, \citenamefont {Broniatowski}, \citenamefont {Camus}, \citenamefont {Cazes}, \citenamefont {Chapellier} \emph {et~al.}}]{arnaud2020edelweiss}%
  \BibitemOpen
  \bibfield  {author} {\bibinfo {author} {\bibfnamefont {Q.}~\bibnamefont {Arnaud}}, \bibinfo {author} {\bibfnamefont {E.}~\bibnamefont {Armengaud}}, \bibinfo {author} {\bibfnamefont {C.}~\bibnamefont {Augier}}, \bibinfo {author} {\bibfnamefont {A.}~\bibnamefont {Beno\^{\i}t}}, \bibinfo {author} {\bibfnamefont {L.}~\bibnamefont {Berg\'e}}, \bibinfo {author} {\bibfnamefont {J.}~\bibnamefont {Billard}}, \bibinfo {author} {\bibfnamefont {A.}~\bibnamefont {Broniatowski}}, \bibinfo {author} {\bibfnamefont {P.}~\bibnamefont {Camus}}, \bibinfo {author} {\bibfnamefont {A.}~\bibnamefont {Cazes}}, \bibinfo {author} {\bibfnamefont {M.}~\bibnamefont {Chapellier}},  \emph {et~al.} (\bibinfo {collaboration} {EDELWEISS Collaboration}),\ }\href {\doibase 10.1103/PhysRevLett.125.141301} {\bibfield  {journal} {\bibinfo  {journal} {Phys. Rev. Lett.}\ }\textbf {\bibinfo {volume} {125}},\ \bibinfo {pages} {141301} (\bibinfo {year} {2020})}\BibitemShut {NoStop}%
\bibitem [{\citenamefont {Conrad}\ and\ \citenamefont {Reimer}(2017)}]{conrad2017indirect}%
  \BibitemOpen
  \bibfield  {author} {\bibinfo {author} {\bibfnamefont {J.}~\bibnamefont {Conrad}}\ and\ \bibinfo {author} {\bibfnamefont {O.}~\bibnamefont {Reimer}},\ }\href {\doibase 10.1038/nphys4049} {\bibfield  {journal} {\bibinfo  {journal} {Nature Physics}\ }\textbf {\bibinfo {volume} {13}},\ \bibinfo {pages} {224} (\bibinfo {year} {2017})}\BibitemShut {NoStop}%
\bibitem [{\citenamefont {Abbasi}\ \emph {et~al.}(2022)\citenamefont {Abbasi}, \citenamefont {Ackermann}, \citenamefont {Adams}, \citenamefont {Aguilar}, \citenamefont {Ahlers}, \citenamefont {Ahrens}, \citenamefont {Alameddine}, \citenamefont {Alispach}, \citenamefont {Alves}, \citenamefont {Amin} \emph {et~al.}}]{Abbasi2022icecube}%
  \BibitemOpen
  \bibfield  {author} {\bibinfo {author} {\bibfnamefont {R.}~\bibnamefont {Abbasi}}, \bibinfo {author} {\bibfnamefont {M.}~\bibnamefont {Ackermann}}, \bibinfo {author} {\bibfnamefont {J.}~\bibnamefont {Adams}}, \bibinfo {author} {\bibfnamefont {J.~A.}\ \bibnamefont {Aguilar}}, \bibinfo {author} {\bibfnamefont {M.}~\bibnamefont {Ahlers}}, \bibinfo {author} {\bibfnamefont {M.}~\bibnamefont {Ahrens}}, \bibinfo {author} {\bibfnamefont {J.~M.}\ \bibnamefont {Alameddine}}, \bibinfo {author} {\bibfnamefont {C.}~\bibnamefont {Alispach}}, \bibinfo {author} {\bibfnamefont {A.~A.}\ \bibnamefont {Alves}}, \bibinfo {author} {\bibfnamefont {N.~M.}\ \bibnamefont {Amin}},  \emph {et~al.} (\bibinfo {collaboration} {IceCube Collaboration}),\ }\href {\doibase 10.1103/PhysRevD.105.062004} {\bibfield  {journal} {\bibinfo  {journal} {Phys. Rev. D}\ }\textbf {\bibinfo {volume} {105}},\ \bibinfo {pages} {062004} (\bibinfo {year} {2022})}\BibitemShut {NoStop}%
\bibitem [{\citenamefont {Meng}\ \emph {et~al.}(2021)\citenamefont {Meng}, \citenamefont {Wang}, \citenamefont {Tao}, \citenamefont {Abdukerim}, \citenamefont {Bo}, \citenamefont {Chen}, \citenamefont {Chen}, \citenamefont {Chen}, \citenamefont {Cheng}, \citenamefont {Cheng} \emph {et~al.}}]{pandax4t2021dm}%
  \BibitemOpen
  \bibfield  {author} {\bibinfo {author} {\bibfnamefont {Y.}~\bibnamefont {Meng}}, \bibinfo {author} {\bibfnamefont {Z.}~\bibnamefont {Wang}}, \bibinfo {author} {\bibfnamefont {Y.}~\bibnamefont {Tao}}, \bibinfo {author} {\bibfnamefont {A.}~\bibnamefont {Abdukerim}}, \bibinfo {author} {\bibfnamefont {Z.}~\bibnamefont {Bo}}, \bibinfo {author} {\bibfnamefont {W.}~\bibnamefont {Chen}}, \bibinfo {author} {\bibfnamefont {X.}~\bibnamefont {Chen}}, \bibinfo {author} {\bibfnamefont {Y.}~\bibnamefont {Chen}}, \bibinfo {author} {\bibfnamefont {C.}~\bibnamefont {Cheng}}, \bibinfo {author} {\bibfnamefont {Y.}~\bibnamefont {Cheng}},  \emph {et~al.} (\bibinfo {collaboration} {PandaX-4T Collaboration}),\ }\href {\doibase 10.1103/PhysRevLett.127.261802} {\bibfield  {journal} {\bibinfo  {journal} {Phys. Rev. Lett.}\ }\textbf {\bibinfo {volume} {127}},\ \bibinfo {pages} {261802} (\bibinfo {year} {2021})}\BibitemShut {NoStop}%
\bibitem [{\citenamefont {Aprile}\ \emph {et~al.}(2022)\citenamefont {Aprile}, \citenamefont {Abe}, \citenamefont {Agostini}, \citenamefont {Ahmed~Maouloud}, \citenamefont {Althueser}, \citenamefont {Andrieu}, \citenamefont {Angelino}, \citenamefont {Angevaare}, \citenamefont {Antochi}, \citenamefont {Ant\'on~Martin} \emph {et~al.}}]{xenonnt2022dm}%
  \BibitemOpen
  \bibfield  {author} {\bibinfo {author} {\bibfnamefont {E.}~\bibnamefont {Aprile}}, \bibinfo {author} {\bibfnamefont {K.}~\bibnamefont {Abe}}, \bibinfo {author} {\bibfnamefont {F.}~\bibnamefont {Agostini}}, \bibinfo {author} {\bibfnamefont {S.}~\bibnamefont {Ahmed~Maouloud}}, \bibinfo {author} {\bibfnamefont {L.}~\bibnamefont {Althueser}}, \bibinfo {author} {\bibfnamefont {B.}~\bibnamefont {Andrieu}}, \bibinfo {author} {\bibfnamefont {E.}~\bibnamefont {Angelino}}, \bibinfo {author} {\bibfnamefont {J.~R.}\ \bibnamefont {Angevaare}}, \bibinfo {author} {\bibfnamefont {V.~C.}\ \bibnamefont {Antochi}}, \bibinfo {author} {\bibfnamefont {D.}~\bibnamefont {Ant\'on~Martin}},  \emph {et~al.} (\bibinfo {collaboration} {XENON Collaboration}),\ }\href {\doibase 10.1103/PhysRevLett.129.161805} {\bibfield  {journal} {\bibinfo  {journal} {Phys. Rev. Lett.}\ }\textbf {\bibinfo {volume} {129}},\ \bibinfo {pages} {161805} (\bibinfo {year} {2022})}\BibitemShut {NoStop}%
\bibitem [{\citenamefont {Battaglieri}\ \emph {et~al.}(2017)\citenamefont {Battaglieri}, \citenamefont {Belloni}, \citenamefont {Chou}, \citenamefont {Cushman}, \citenamefont {Echenard}, \citenamefont {Essig}, \citenamefont {Estrada}, \citenamefont {Feng}, \citenamefont {Flaugher}, \citenamefont {Fox} \emph {et~al.}}]{battaglieri2017us}%
  \BibitemOpen
  \bibfield  {author} {\bibinfo {author} {\bibfnamefont {M.}~\bibnamefont {Battaglieri}}, \bibinfo {author} {\bibfnamefont {A.}~\bibnamefont {Belloni}}, \bibinfo {author} {\bibfnamefont {A.}~\bibnamefont {Chou}}, \bibinfo {author} {\bibfnamefont {P.}~\bibnamefont {Cushman}}, \bibinfo {author} {\bibfnamefont {B.}~\bibnamefont {Echenard}}, \bibinfo {author} {\bibfnamefont {R.}~\bibnamefont {Essig}}, \bibinfo {author} {\bibfnamefont {J.}~\bibnamefont {Estrada}}, \bibinfo {author} {\bibfnamefont {J.~L.}\ \bibnamefont {Feng}}, \bibinfo {author} {\bibfnamefont {B.}~\bibnamefont {Flaugher}}, \bibinfo {author} {\bibfnamefont {P.~J.}\ \bibnamefont {Fox}},  \emph {et~al.},\ }\href {\doibase 10.48550/arXiv.1707.04591} {\bibfield  {journal} {\bibinfo  {journal} {arXiv preprint arXiv:1707.04591}\ } (\bibinfo {year} {2017}),\ 10.48550/arXiv.1707.04591}\BibitemShut {NoStop}%
\bibitem [{\citenamefont {Boveia}\ \emph {et~al.}(2022)\citenamefont {Boveia}, \citenamefont {Chen}, \citenamefont {Doglioni}, \citenamefont {Drlica-Wagner}, \citenamefont {Gori}, \citenamefont {Lippincott}, \citenamefont {Monzani}, \citenamefont {Prescod-Weinstein}, \citenamefont {Shakya}, \citenamefont {Slatyer} \emph {et~al.}}]{boveia2022snowmass}%
  \BibitemOpen
  \bibfield  {author} {\bibinfo {author} {\bibfnamefont {A.}~\bibnamefont {Boveia}}, \bibinfo {author} {\bibfnamefont {T.~Y.}\ \bibnamefont {Chen}}, \bibinfo {author} {\bibfnamefont {C.}~\bibnamefont {Doglioni}}, \bibinfo {author} {\bibfnamefont {A.}~\bibnamefont {Drlica-Wagner}}, \bibinfo {author} {\bibfnamefont {S.}~\bibnamefont {Gori}}, \bibinfo {author} {\bibfnamefont {W.~H.}\ \bibnamefont {Lippincott}}, \bibinfo {author} {\bibfnamefont {M.~E.}\ \bibnamefont {Monzani}}, \bibinfo {author} {\bibfnamefont {C.}~\bibnamefont {Prescod-Weinstein}}, \bibinfo {author} {\bibfnamefont {B.}~\bibnamefont {Shakya}}, \bibinfo {author} {\bibfnamefont {T.~R.}\ \bibnamefont {Slatyer}},  \emph {et~al.},\ }\href {\doibase 10.48550/arXiv.2210.01770} {\bibfield  {journal} {\bibinfo  {journal} {arXiv preprint arXiv:2210.01770}\ } (\bibinfo {year} {2022}),\ 10.48550/arXiv.2210.01770}\BibitemShut {NoStop}%
\bibitem [{\citenamefont {Group}\ \emph {et~al.}(2022)\citenamefont {Group} \emph {et~al.}}]{particle2022reviewDM}%
  \BibitemOpen
  \bibfield  {author} {\bibinfo {author} {\bibfnamefont {P.~D.}\ \bibnamefont {Group}} \emph {et~al.},\ }\href@noop {} {\bibfield  {journal} {\bibinfo  {journal} {Progress of Theoretical and Experimental Physics}\ }\textbf {\bibinfo {volume} {2022}} (\bibinfo {year} {2022})}\BibitemShut {NoStop}%
\bibitem [{\citenamefont {Essig}\ \emph {et~al.}(2012)\citenamefont {Essig}, \citenamefont {Mardon},\ and\ \citenamefont {Volansky}}]{essig2012subgev}%
  \BibitemOpen
  \bibfield  {author} {\bibinfo {author} {\bibfnamefont {R.}~\bibnamefont {Essig}}, \bibinfo {author} {\bibfnamefont {J.}~\bibnamefont {Mardon}}, \ and\ \bibinfo {author} {\bibfnamefont {T.}~\bibnamefont {Volansky}},\ }\href {\doibase 10.1103/PhysRevD.85.076007} {\bibfield  {journal} {\bibinfo  {journal} {Phys. Rev. D}\ }\textbf {\bibinfo {volume} {85}},\ \bibinfo {pages} {076007} (\bibinfo {year} {2012})}\BibitemShut {NoStop}%
\bibitem [{\citenamefont {Graham}\ \emph {et~al.}(2012)\citenamefont {Graham}, \citenamefont {Kaplan}, \citenamefont {Rajendran},\ and\ \citenamefont {Walters}}]{graham2012subgev}%
  \BibitemOpen
  \bibfield  {author} {\bibinfo {author} {\bibfnamefont {P.~W.}\ \bibnamefont {Graham}}, \bibinfo {author} {\bibfnamefont {D.~E.}\ \bibnamefont {Kaplan}}, \bibinfo {author} {\bibfnamefont {S.}~\bibnamefont {Rajendran}}, \ and\ \bibinfo {author} {\bibfnamefont {M.~T.}\ \bibnamefont {Walters}},\ }\href {\doibase https://doi.org/10.1016/j.dark.2012.09.001} {\bibfield  {journal} {\bibinfo  {journal} {Physics of the Dark Universe}\ }\textbf {\bibinfo {volume} {1}},\ \bibinfo {pages} {32} (\bibinfo {year} {2012})},\ \bibinfo {note} {next Decade in Dark Matter and Dark Energy}\BibitemShut {NoStop}%
\bibitem [{\citenamefont {Essig}\ \emph {et~al.}(2016)\citenamefont {Essig}, \citenamefont {Fernandez-Serra}, \citenamefont {Mardon}, \citenamefont {Soto}, \citenamefont {Volansky},\ and\ \citenamefont {Yu}}]{essig2016direct}%
  \BibitemOpen
  \bibfield  {author} {\bibinfo {author} {\bibfnamefont {R.}~\bibnamefont {Essig}}, \bibinfo {author} {\bibfnamefont {M.}~\bibnamefont {Fernandez-Serra}}, \bibinfo {author} {\bibfnamefont {J.}~\bibnamefont {Mardon}}, \bibinfo {author} {\bibfnamefont {A.}~\bibnamefont {Soto}}, \bibinfo {author} {\bibfnamefont {T.}~\bibnamefont {Volansky}}, \ and\ \bibinfo {author} {\bibfnamefont {T.-T.}\ \bibnamefont {Yu}},\ }\href {\doibase 10.1007/JHEP05(2016)046} {\bibfield  {journal} {\bibinfo  {journal} {Journal of High Energy Physics}\ }\textbf {\bibinfo {volume} {2016}},\ \bibinfo {pages} {1} (\bibinfo {year} {2016})}\BibitemShut {NoStop}%
\bibitem [{\citenamefont {Pospelov}\ \emph {et~al.}(2008)\citenamefont {Pospelov}, \citenamefont {Ritz},\ and\ \citenamefont {Voloshin}}]{pospelov2008superwimp}%
  \BibitemOpen
  \bibfield  {author} {\bibinfo {author} {\bibfnamefont {M.}~\bibnamefont {Pospelov}}, \bibinfo {author} {\bibfnamefont {A.}~\bibnamefont {Ritz}}, \ and\ \bibinfo {author} {\bibfnamefont {M.}~\bibnamefont {Voloshin}},\ }\href {\doibase 10.1103/PhysRevD.78.115012} {\bibfield  {journal} {\bibinfo  {journal} {Phys. Rev. D}\ }\textbf {\bibinfo {volume} {78}},\ \bibinfo {pages} {115012} (\bibinfo {year} {2008})}\BibitemShut {NoStop}%
\bibitem [{\citenamefont {Agostini}\ \emph {et~al.}(2020)\citenamefont {Agostini}, \citenamefont {Bakalyarov}, \citenamefont {Balata}, \citenamefont {Barabanov}, \citenamefont {Baudis}, \citenamefont {Bauer}, \citenamefont {Bellotti}, \citenamefont {Belogurov}, \citenamefont {Bettini}, \citenamefont {Bezrukov} \emph {et~al.}}]{gerda2020superwimp}%
  \BibitemOpen
  \bibfield  {author} {\bibinfo {author} {\bibfnamefont {M.}~\bibnamefont {Agostini}}, \bibinfo {author} {\bibfnamefont {A.~M.}\ \bibnamefont {Bakalyarov}}, \bibinfo {author} {\bibfnamefont {M.}~\bibnamefont {Balata}}, \bibinfo {author} {\bibfnamefont {I.}~\bibnamefont {Barabanov}}, \bibinfo {author} {\bibfnamefont {L.}~\bibnamefont {Baudis}}, \bibinfo {author} {\bibfnamefont {C.}~\bibnamefont {Bauer}}, \bibinfo {author} {\bibfnamefont {E.}~\bibnamefont {Bellotti}}, \bibinfo {author} {\bibfnamefont {S.}~\bibnamefont {Belogurov}}, \bibinfo {author} {\bibfnamefont {A.}~\bibnamefont {Bettini}}, \bibinfo {author} {\bibfnamefont {L.}~\bibnamefont {Bezrukov}},  \emph {et~al.} (\bibinfo {collaboration} {GERDA Collaboration}),\ }\href {\doibase 10.1103/PhysRevLett.125.011801} {\bibfield  {journal} {\bibinfo  {journal} {Phys. Rev. Lett.}\ }\textbf {\bibinfo {volume} {125}},\ \bibinfo {pages} {011801} (\bibinfo {year} {2020})}\BibitemShut {NoStop}%
\bibitem [{\citenamefont {Arnquist}\ \emph {et~al.}(2024)\citenamefont {Arnquist}, \citenamefont {Avignone}, \citenamefont {Barabash}, \citenamefont {Barton}, \citenamefont {Bhimani}, \citenamefont {Blalock}, \citenamefont {Bos}, \citenamefont {Busch}, \citenamefont {Buuck}, \citenamefont {Caldwell} \emph {et~al.}}]{arnquist2022exotic}%
  \BibitemOpen
  \bibfield  {author} {\bibinfo {author} {\bibfnamefont {I.~J.}\ \bibnamefont {Arnquist}}, \bibinfo {author} {\bibfnamefont {F.~T.}\ \bibnamefont {Avignone}}, \bibinfo {author} {\bibfnamefont {A.~S.}\ \bibnamefont {Barabash}}, \bibinfo {author} {\bibfnamefont {C.~J.}\ \bibnamefont {Barton}}, \bibinfo {author} {\bibfnamefont {K.~H.}\ \bibnamefont {Bhimani}}, \bibinfo {author} {\bibfnamefont {E.}~\bibnamefont {Blalock}}, \bibinfo {author} {\bibfnamefont {B.}~\bibnamefont {Bos}}, \bibinfo {author} {\bibfnamefont {M.}~\bibnamefont {Busch}}, \bibinfo {author} {\bibfnamefont {M.}~\bibnamefont {Buuck}}, \bibinfo {author} {\bibfnamefont {T.~S.}\ \bibnamefont {Caldwell}},  \emph {et~al.} (\bibinfo {collaboration} {Majorana Collaboration}),\ }\href {\doibase 10.1103/PhysRevLett.132.041001} {\bibfield  {journal} {\bibinfo  {journal} {Phys. Rev. Lett.}\ }\textbf {\bibinfo {volume} {132}},\ \bibinfo {pages} {041001} (\bibinfo {year} {2024})}\BibitemShut {NoStop}%
\bibitem [{\citenamefont {Vergados}\ and\ \citenamefont {Ejiri}(2005)}]{vergados2005migdal}%
  \BibitemOpen
  \bibfield  {author} {\bibinfo {author} {\bibfnamefont {J.}~\bibnamefont {Vergados}}\ and\ \bibinfo {author} {\bibfnamefont {H.}~\bibnamefont {Ejiri}},\ }\href {\doibase https://doi.org/10.1016/j.physletb.2004.11.085} {\bibfield  {journal} {\bibinfo  {journal} {Physics Letters B}\ }\textbf {\bibinfo {volume} {606}},\ \bibinfo {pages} {313} (\bibinfo {year} {2005})}\BibitemShut {NoStop}%
\bibitem [{\citenamefont {Bernabei}\ \emph {et~al.}(2007)\citenamefont {Bernabei}, \citenamefont {Belli}, \citenamefont {Montecchia}, \citenamefont {Nozzoli}, \citenamefont {Cappella}, \citenamefont {Incicchitti}, \citenamefont {Prosperi}, \citenamefont {Cerulli}, \citenamefont {Dai}, \citenamefont {He} \emph {et~al.}}]{bernabei2007migdal}%
  \BibitemOpen
  \bibfield  {author} {\bibinfo {author} {\bibfnamefont {R.}~\bibnamefont {Bernabei}}, \bibinfo {author} {\bibfnamefont {P.}~\bibnamefont {Belli}}, \bibinfo {author} {\bibfnamefont {F.}~\bibnamefont {Montecchia}}, \bibinfo {author} {\bibfnamefont {F.}~\bibnamefont {Nozzoli}}, \bibinfo {author} {\bibfnamefont {F.}~\bibnamefont {Cappella}}, \bibinfo {author} {\bibfnamefont {A.}~\bibnamefont {Incicchitti}}, \bibinfo {author} {\bibfnamefont {D.}~\bibnamefont {Prosperi}}, \bibinfo {author} {\bibfnamefont {R.}~\bibnamefont {Cerulli}}, \bibinfo {author} {\bibfnamefont {C.}~\bibnamefont {Dai}}, \bibinfo {author} {\bibfnamefont {H.}~\bibnamefont {He}},  \emph {et~al.},\ }\href {\doibase 10.1142/S0217751X07037093} {\bibfield  {journal} {\bibinfo  {journal} {International Journal of Modern Physics A}\ }\textbf {\bibinfo {volume} {22}},\ \bibinfo {pages} {3155} (\bibinfo {year} {2007})}\BibitemShut {NoStop}%
\bibitem [{\citenamefont {Ibe}\ \emph {et~al.}(2018)\citenamefont {Ibe}, \citenamefont {Nakano}, \citenamefont {Shoji},\ and\ \citenamefont {Suzuki}}]{ibe2018migdal}%
  \BibitemOpen
  \bibfield  {author} {\bibinfo {author} {\bibfnamefont {M.}~\bibnamefont {Ibe}}, \bibinfo {author} {\bibfnamefont {W.}~\bibnamefont {Nakano}}, \bibinfo {author} {\bibfnamefont {Y.}~\bibnamefont {Shoji}}, \ and\ \bibinfo {author} {\bibfnamefont {K.}~\bibnamefont {Suzuki}},\ }\href {\doibase 10.1007/JHEP03(2018)194} {\bibfield  {journal} {\bibinfo  {journal} {Journal of High Energy Physics}\ }\textbf {\bibinfo {volume} {2018}} (\bibinfo {year} {2018}),\ 10.1007/JHEP03(2018)194}\BibitemShut {NoStop}%
\bibitem [{\citenamefont {Essig}\ \emph {et~al.}(2020)\citenamefont {Essig}, \citenamefont {Pradler}, \citenamefont {Sholapurkar},\ and\ \citenamefont {Yu}}]{essig2020migdal}%
  \BibitemOpen
  \bibfield  {author} {\bibinfo {author} {\bibfnamefont {R.}~\bibnamefont {Essig}}, \bibinfo {author} {\bibfnamefont {J.}~\bibnamefont {Pradler}}, \bibinfo {author} {\bibfnamefont {M.}~\bibnamefont {Sholapurkar}}, \ and\ \bibinfo {author} {\bibfnamefont {T.-T.}\ \bibnamefont {Yu}},\ }\href {\doibase 10.1103/PhysRevLett.124.021801} {\bibfield  {journal} {\bibinfo  {journal} {Phys. Rev. Lett.}\ }\textbf {\bibinfo {volume} {124}},\ \bibinfo {pages} {021801} (\bibinfo {year} {2020})}\BibitemShut {NoStop}%
\bibitem [{\citenamefont {Tiffenberg}\ \emph {et~al.}(2017)\citenamefont {Tiffenberg}, \citenamefont {Sofo-Haro}, \citenamefont {Drlica-Wagner}, \citenamefont {Essig}, \citenamefont {Guardincerri}, \citenamefont {Holland}, \citenamefont {Volansky},\ and\ \citenamefont {Yu}}]{tiffenberg2017sccd}%
  \BibitemOpen
  \bibfield  {author} {\bibinfo {author} {\bibfnamefont {J.}~\bibnamefont {Tiffenberg}}, \bibinfo {author} {\bibfnamefont {M.}~\bibnamefont {Sofo-Haro}}, \bibinfo {author} {\bibfnamefont {A.}~\bibnamefont {Drlica-Wagner}}, \bibinfo {author} {\bibfnamefont {R.}~\bibnamefont {Essig}}, \bibinfo {author} {\bibfnamefont {Y.}~\bibnamefont {Guardincerri}}, \bibinfo {author} {\bibfnamefont {S.}~\bibnamefont {Holland}}, \bibinfo {author} {\bibfnamefont {T.}~\bibnamefont {Volansky}}, \ and\ \bibinfo {author} {\bibfnamefont {T.-T.}\ \bibnamefont {Yu}},\ }\href {\doibase 10.1103/PhysRevLett.119.131802} {\bibfield  {journal} {\bibinfo  {journal} {Phys. Rev. Lett.}\ }\textbf {\bibinfo {volume} {119}},\ \bibinfo {pages} {131802} (\bibinfo {year} {2017})}\BibitemShut {NoStop}%
\bibitem [{\citenamefont {Fichet}(2018)}]{fichet2018quantumdm}%
  \BibitemOpen
  \bibfield  {author} {\bibinfo {author} {\bibfnamefont {S.}~\bibnamefont {Fichet}},\ }\href {\doibase 10.1103/PhysRevLett.120.131801} {\bibfield  {journal} {\bibinfo  {journal} {Phys. Rev. Lett.}\ }\textbf {\bibinfo {volume} {120}},\ \bibinfo {pages} {131801} (\bibinfo {year} {2018})}\BibitemShut {NoStop}%
\bibitem [{\citenamefont {Guo}\ and\ \citenamefont {McKinsey}(2013)}]{guo2013superfluid}%
  \BibitemOpen
  \bibfield  {author} {\bibinfo {author} {\bibfnamefont {W.}~\bibnamefont {Guo}}\ and\ \bibinfo {author} {\bibfnamefont {D.~N.}\ \bibnamefont {McKinsey}},\ }\href {\doibase 10.1103/PhysRevD.87.115001} {\bibfield  {journal} {\bibinfo  {journal} {Phys. Rev. D}\ }\textbf {\bibinfo {volume} {87}},\ \bibinfo {pages} {115001} (\bibinfo {year} {2013})}\BibitemShut {NoStop}%
\bibitem [{\citenamefont {Caputo}\ \emph {et~al.}(2020)\citenamefont {Caputo}, \citenamefont {Esposito}, \citenamefont {Geoffray}, \citenamefont {Polosa},\ and\ \citenamefont {Sun}}]{caputo2020superfluid}%
  \BibitemOpen
  \bibfield  {author} {\bibinfo {author} {\bibfnamefont {A.}~\bibnamefont {Caputo}}, \bibinfo {author} {\bibfnamefont {A.}~\bibnamefont {Esposito}}, \bibinfo {author} {\bibfnamefont {E.}~\bibnamefont {Geoffray}}, \bibinfo {author} {\bibfnamefont {A.~D.}\ \bibnamefont {Polosa}}, \ and\ \bibinfo {author} {\bibfnamefont {S.}~\bibnamefont {Sun}},\ }\href {\doibase https://doi.org/10.1016/j.physletb.2020.135258} {\bibfield  {journal} {\bibinfo  {journal} {Physics Letters B}\ }\textbf {\bibinfo {volume} {802}},\ \bibinfo {pages} {135258} (\bibinfo {year} {2020})}\BibitemShut {NoStop}%
\bibitem [{\citenamefont {Maris}\ \emph {et~al.}(2017)\citenamefont {Maris}, \citenamefont {Seidel},\ and\ \citenamefont {Stein}}]{maris2017evaporation}%
  \BibitemOpen
  \bibfield  {author} {\bibinfo {author} {\bibfnamefont {H.~J.}\ \bibnamefont {Maris}}, \bibinfo {author} {\bibfnamefont {G.~M.}\ \bibnamefont {Seidel}}, \ and\ \bibinfo {author} {\bibfnamefont {D.}~\bibnamefont {Stein}},\ }\href {\doibase 10.1103/PhysRevLett.119.181303} {\bibfield  {journal} {\bibinfo  {journal} {Phys. Rev. Lett.}\ }\textbf {\bibinfo {volume} {119}},\ \bibinfo {pages} {181303} (\bibinfo {year} {2017})}\BibitemShut {NoStop}%
\bibitem [{\citenamefont {Hochberg}\ \emph {et~al.}(2017)\citenamefont {Hochberg}, \citenamefont {Kahn}, \citenamefont {Lisanti}, \citenamefont {Tully},\ and\ \citenamefont {Zurek}}]{hochberg2017grephene}%
  \BibitemOpen
  \bibfield  {author} {\bibinfo {author} {\bibfnamefont {Y.}~\bibnamefont {Hochberg}}, \bibinfo {author} {\bibfnamefont {Y.}~\bibnamefont {Kahn}}, \bibinfo {author} {\bibfnamefont {M.}~\bibnamefont {Lisanti}}, \bibinfo {author} {\bibfnamefont {C.~G.}\ \bibnamefont {Tully}}, \ and\ \bibinfo {author} {\bibfnamefont {K.~M.}\ \bibnamefont {Zurek}},\ }\href {\doibase https://doi.org/10.1016/j.physletb.2017.06.051} {\bibfield  {journal} {\bibinfo  {journal} {Physics Letters B}\ }\textbf {\bibinfo {volume} {772}},\ \bibinfo {pages} {239} (\bibinfo {year} {2017})}\BibitemShut {NoStop}%
\bibitem [{\citenamefont {Kim}\ \emph {et~al.}(2020)\citenamefont {Kim}, \citenamefont {Park}, \citenamefont {Fong},\ and\ \citenamefont {Lee}}]{kim2020detection}%
  \BibitemOpen
  \bibfield  {author} {\bibinfo {author} {\bibfnamefont {D.}~\bibnamefont {Kim}}, \bibinfo {author} {\bibfnamefont {J.-C.}\ \bibnamefont {Park}}, \bibinfo {author} {\bibfnamefont {K.~C.}\ \bibnamefont {Fong}}, \ and\ \bibinfo {author} {\bibfnamefont {G.-H.}\ \bibnamefont {Lee}},\ }\href {\doibase 10.48550/arXiv.2002.07821} {\bibfield  {journal} {\bibinfo  {journal} {arXiv preprint arXiv:2002.07821}\ } (\bibinfo {year} {2020}),\ 10.48550/arXiv.2002.07821}\BibitemShut {NoStop}%
\bibitem [{\citenamefont {Hochberg}\ \emph {et~al.}(2019)\citenamefont {Hochberg}, \citenamefont {Charaev}, \citenamefont {Nam}, \citenamefont {Verma}, \citenamefont {Colangelo},\ and\ \citenamefont {Berggren}}]{hochberg2019nanowire}%
  \BibitemOpen
  \bibfield  {author} {\bibinfo {author} {\bibfnamefont {Y.}~\bibnamefont {Hochberg}}, \bibinfo {author} {\bibfnamefont {I.}~\bibnamefont {Charaev}}, \bibinfo {author} {\bibfnamefont {S.-W.}\ \bibnamefont {Nam}}, \bibinfo {author} {\bibfnamefont {V.}~\bibnamefont {Verma}}, \bibinfo {author} {\bibfnamefont {M.}~\bibnamefont {Colangelo}}, \ and\ \bibinfo {author} {\bibfnamefont {K.~K.}\ \bibnamefont {Berggren}},\ }\href {\doibase 10.1103/PhysRevLett.123.151802} {\bibfield  {journal} {\bibinfo  {journal} {Phys. Rev. Lett.}\ }\textbf {\bibinfo {volume} {123}},\ \bibinfo {pages} {151802} (\bibinfo {year} {2019})}\BibitemShut {NoStop}%
\bibitem [{\citenamefont {Carter}\ \emph {et~al.}(2017)\citenamefont {Carter}, \citenamefont {Hertel}, \citenamefont {Rooks}, \citenamefont {McClintock}, \citenamefont {McKinsey},\ and\ \citenamefont {Prober}}]{carter2017calorimetric}%
  \BibitemOpen
  \bibfield  {author} {\bibinfo {author} {\bibfnamefont {F.~W.}\ \bibnamefont {Carter}}, \bibinfo {author} {\bibfnamefont {S.~A.}\ \bibnamefont {Hertel}}, \bibinfo {author} {\bibfnamefont {M.}~\bibnamefont {Rooks}}, \bibinfo {author} {\bibfnamefont {P.}~\bibnamefont {McClintock}}, \bibinfo {author} {\bibfnamefont {D.}~\bibnamefont {McKinsey}}, \ and\ \bibinfo {author} {\bibfnamefont {D.}~\bibnamefont {Prober}},\ }\href {\doibase 10.1007/s10909-016-1666-x} {\bibfield  {journal} {\bibinfo  {journal} {Journal of low temperature physics}\ }\textbf {\bibinfo {volume} {186}},\ \bibinfo {pages} {183} (\bibinfo {year} {2017})}\BibitemShut {NoStop}%
\bibitem [{\citenamefont {Essig}\ \emph {et~al.}(2017)\citenamefont {Essig}, \citenamefont {Mardon}, \citenamefont {Slone},\ and\ \citenamefont {Volansky}}]{essig2017chemical}%
  \BibitemOpen
  \bibfield  {author} {\bibinfo {author} {\bibfnamefont {R.}~\bibnamefont {Essig}}, \bibinfo {author} {\bibfnamefont {J.}~\bibnamefont {Mardon}}, \bibinfo {author} {\bibfnamefont {O.}~\bibnamefont {Slone}}, \ and\ \bibinfo {author} {\bibfnamefont {T.}~\bibnamefont {Volansky}},\ }\href {\doibase 10.1103/PhysRevD.95.056011} {\bibfield  {journal} {\bibinfo  {journal} {Phys. Rev. D}\ }\textbf {\bibinfo {volume} {95}},\ \bibinfo {pages} {056011} (\bibinfo {year} {2017})}\BibitemShut {NoStop}%
\bibitem [{\citenamefont {Hochberg}\ \emph {et~al.}(2016{\natexlab{a}})\citenamefont {Hochberg}, \citenamefont {Zhao},\ and\ \citenamefont {Zurek}}]{hochberg2016superconductor}%
  \BibitemOpen
  \bibfield  {author} {\bibinfo {author} {\bibfnamefont {Y.}~\bibnamefont {Hochberg}}, \bibinfo {author} {\bibfnamefont {Y.}~\bibnamefont {Zhao}}, \ and\ \bibinfo {author} {\bibfnamefont {K.~M.}\ \bibnamefont {Zurek}},\ }\href {\doibase 10.1103/PhysRevLett.116.011301} {\bibfield  {journal} {\bibinfo  {journal} {Phys. Rev. Lett.}\ }\textbf {\bibinfo {volume} {116}},\ \bibinfo {pages} {011301} (\bibinfo {year} {2016}{\natexlab{a}})}\BibitemShut {NoStop}%
\bibitem [{\citenamefont {Hochberg}\ \emph {et~al.}(2018)\citenamefont {Hochberg}, \citenamefont {Kahn}, \citenamefont {Lisanti}, \citenamefont {Zurek}, \citenamefont {Grushin}, \citenamefont {Ilan}, \citenamefont {Griffin}, \citenamefont {Liu}, \citenamefont {Weber},\ and\ \citenamefont {Neaton}}]{hochberg2018dirac}%
  \BibitemOpen
  \bibfield  {author} {\bibinfo {author} {\bibfnamefont {Y.}~\bibnamefont {Hochberg}}, \bibinfo {author} {\bibfnamefont {Y.}~\bibnamefont {Kahn}}, \bibinfo {author} {\bibfnamefont {M.}~\bibnamefont {Lisanti}}, \bibinfo {author} {\bibfnamefont {K.~M.}\ \bibnamefont {Zurek}}, \bibinfo {author} {\bibfnamefont {A.~G.}\ \bibnamefont {Grushin}}, \bibinfo {author} {\bibfnamefont {R.}~\bibnamefont {Ilan}}, \bibinfo {author} {\bibfnamefont {S.~M.}\ \bibnamefont {Griffin}}, \bibinfo {author} {\bibfnamefont {Z.-F.}\ \bibnamefont {Liu}}, \bibinfo {author} {\bibfnamefont {S.~F.}\ \bibnamefont {Weber}}, \ and\ \bibinfo {author} {\bibfnamefont {J.~B.}\ \bibnamefont {Neaton}},\ }\href {\doibase 10.1103/PhysRevD.97.015004} {\bibfield  {journal} {\bibinfo  {journal} {Phys. Rev. D}\ }\textbf {\bibinfo {volume} {97}},\ \bibinfo {pages} {015004} (\bibinfo {year} {2018})}\BibitemShut {NoStop}%
\bibitem [{\citenamefont {Hochberg}\ \emph {et~al.}(2016{\natexlab{b}})\citenamefont {Hochberg}, \citenamefont {Pyle}, \citenamefont {Zhao},\ and\ \citenamefont {Zurek}}]{hochberg2016detecting}%
  \BibitemOpen
  \bibfield  {author} {\bibinfo {author} {\bibfnamefont {Y.}~\bibnamefont {Hochberg}}, \bibinfo {author} {\bibfnamefont {M.}~\bibnamefont {Pyle}}, \bibinfo {author} {\bibfnamefont {Y.}~\bibnamefont {Zhao}}, \ and\ \bibinfo {author} {\bibfnamefont {K.~M.}\ \bibnamefont {Zurek}},\ }\href {\doibase 10.1007/JHEP08(2016)057} {\bibfield  {journal} {\bibinfo  {journal} {Journal of High Energy Physics}\ }\textbf {\bibinfo {volume} {2016}},\ \bibinfo {pages} {1} (\bibinfo {year} {2016}{\natexlab{b}})}\BibitemShut {NoStop}%
\bibitem [{\citenamefont {Kurinsky}\ \emph {et~al.}(2019)\citenamefont {Kurinsky}, \citenamefont {Yu}, \citenamefont {Hochberg},\ and\ \citenamefont {Cabrera}}]{kurinsky2019diamond}%
  \BibitemOpen
  \bibfield  {author} {\bibinfo {author} {\bibfnamefont {N.}~\bibnamefont {Kurinsky}}, \bibinfo {author} {\bibfnamefont {T.~C.}\ \bibnamefont {Yu}}, \bibinfo {author} {\bibfnamefont {Y.}~\bibnamefont {Hochberg}}, \ and\ \bibinfo {author} {\bibfnamefont {B.}~\bibnamefont {Cabrera}},\ }\href {\doibase 10.1103/PhysRevD.99.123005} {\bibfield  {journal} {\bibinfo  {journal} {Phys. Rev. D}\ }\textbf {\bibinfo {volume} {99}},\ \bibinfo {pages} {123005} (\bibinfo {year} {2019})}\BibitemShut {NoStop}%
\bibitem [{\citenamefont {Canonica}\ \emph {et~al.}(2020)\citenamefont {Canonica}, \citenamefont {Abdelhameed}, \citenamefont {Bauer}, \citenamefont {Bento}, \citenamefont {Bertoldo}, \citenamefont {Ferreiro~Iachellini}, \citenamefont {Fuchs}, \citenamefont {Hauff}, \citenamefont {Mancuso}, \citenamefont {Petricca} \emph {et~al.}}]{canonica2020diamond}%
  \BibitemOpen
  \bibfield  {author} {\bibinfo {author} {\bibfnamefont {L.}~\bibnamefont {Canonica}}, \bibinfo {author} {\bibfnamefont {A.}~\bibnamefont {Abdelhameed}}, \bibinfo {author} {\bibfnamefont {P.}~\bibnamefont {Bauer}}, \bibinfo {author} {\bibfnamefont {A.}~\bibnamefont {Bento}}, \bibinfo {author} {\bibfnamefont {E.}~\bibnamefont {Bertoldo}}, \bibinfo {author} {\bibfnamefont {N.}~\bibnamefont {Ferreiro~Iachellini}}, \bibinfo {author} {\bibfnamefont {D.}~\bibnamefont {Fuchs}}, \bibinfo {author} {\bibfnamefont {D.}~\bibnamefont {Hauff}}, \bibinfo {author} {\bibfnamefont {M.}~\bibnamefont {Mancuso}}, \bibinfo {author} {\bibfnamefont {F.}~\bibnamefont {Petricca}},  \emph {et~al.},\ }\href {\doibase 10.1007/s10909-020-02350-4} {\bibfield  {journal} {\bibinfo  {journal} {Journal of Low Temperature Physics}\ }\textbf {\bibinfo {volume} {199}},\ \bibinfo {pages} {606} (\bibinfo {year} {2020})}\BibitemShut {NoStop}%
\bibitem [{\citenamefont {Marshall}\ \emph {et~al.}(2021)\citenamefont {Marshall}, \citenamefont {Turner}, \citenamefont {Ku}, \citenamefont {Phillips},\ and\ \citenamefont {Walsworth}}]{marshall2021diamond}%
  \BibitemOpen
  \bibfield  {author} {\bibinfo {author} {\bibfnamefont {M.~C.}\ \bibnamefont {Marshall}}, \bibinfo {author} {\bibfnamefont {M.~J.}\ \bibnamefont {Turner}}, \bibinfo {author} {\bibfnamefont {M.~J.~H.}\ \bibnamefont {Ku}}, \bibinfo {author} {\bibfnamefont {D.~F.}\ \bibnamefont {Phillips}}, \ and\ \bibinfo {author} {\bibfnamefont {R.~L.}\ \bibnamefont {Walsworth}},\ }\href {\doibase 10.1088/2058-9565/abe5ed} {\bibfield  {journal} {\bibinfo  {journal} {Quantum Science and Technology}\ }\textbf {\bibinfo {volume} {6}},\ \bibinfo {pages} {024011} (\bibinfo {year} {2021})}\BibitemShut {NoStop}%
\bibitem [{\citenamefont {Abdelhameed}\ \emph {et~al.}(2022)\citenamefont {Abdelhameed}, \citenamefont {Angloher}, \citenamefont {Bento}, \citenamefont {Bertoldo}, \citenamefont {Bertolini}, \citenamefont {Canonica}, \citenamefont {Iachellini}, \citenamefont {Fuchs}, \citenamefont {Garai}, \citenamefont {Hauff} \emph {et~al.}}]{abdelhameed2022diamond}%
  \BibitemOpen
  \bibfield  {author} {\bibinfo {author} {\bibfnamefont {A.}~\bibnamefont {Abdelhameed}}, \bibinfo {author} {\bibfnamefont {G.}~\bibnamefont {Angloher}}, \bibinfo {author} {\bibfnamefont {A.}~\bibnamefont {Bento}}, \bibinfo {author} {\bibfnamefont {E.}~\bibnamefont {Bertoldo}}, \bibinfo {author} {\bibfnamefont {A.}~\bibnamefont {Bertolini}}, \bibinfo {author} {\bibfnamefont {L.}~\bibnamefont {Canonica}}, \bibinfo {author} {\bibfnamefont {N.~F.}\ \bibnamefont {Iachellini}}, \bibinfo {author} {\bibfnamefont {D.}~\bibnamefont {Fuchs}}, \bibinfo {author} {\bibfnamefont {A.}~\bibnamefont {Garai}}, \bibinfo {author} {\bibfnamefont {D.}~\bibnamefont {Hauff}},  \emph {et~al.},\ }\href {\doibase 10.1140/epjc/s10052-022-10829-5} {\bibfield  {journal} {\bibinfo  {journal} {The European Physical Journal C}\ }\textbf {\bibinfo {volume} {82}},\ \bibinfo {pages} {851} (\bibinfo {year} {2022})}\BibitemShut {NoStop}%
\bibitem [{\citenamefont {Kempf}\ \emph {et~al.}(2018)\citenamefont {Kempf}, \citenamefont {Fleischmann}, \citenamefont {Gastaldo},\ and\ \citenamefont {Enss}}]{kempf2018physics}%
  \BibitemOpen
  \bibfield  {author} {\bibinfo {author} {\bibfnamefont {S.}~\bibnamefont {Kempf}}, \bibinfo {author} {\bibfnamefont {A.}~\bibnamefont {Fleischmann}}, \bibinfo {author} {\bibfnamefont {L.}~\bibnamefont {Gastaldo}}, \ and\ \bibinfo {author} {\bibfnamefont {C.}~\bibnamefont {Enss}},\ }\href {\doibase 10.1007/s10909-018-1891-6} {\bibfield  {journal} {\bibinfo  {journal} {Journal of Low Temperature Physics}\ }\textbf {\bibinfo {volume} {193}},\ \bibinfo {pages} {365} (\bibinfo {year} {2018})}\BibitemShut {NoStop}%
\bibitem [{\citenamefont {Fleischmann}\ \emph {et~al.}(2005)\citenamefont {Fleischmann}, \citenamefont {Enss},\ and\ \citenamefont {Seidel}}]{Fleischmann2005}%
  \BibitemOpen
  \bibfield  {author} {\bibinfo {author} {\bibfnamefont {A.}~\bibnamefont {Fleischmann}}, \bibinfo {author} {\bibfnamefont {C.}~\bibnamefont {Enss}}, \ and\ \bibinfo {author} {\bibfnamefont {G.}~\bibnamefont {Seidel}},\ }\enquote {\bibinfo {title} {Metallic magnetic calorimeters},}\ in\ \href {\doibase 10.1007/10933596_4} {\emph {\bibinfo {booktitle} {Cryogenic Particle Detection}}},\ \bibinfo {editor} {edited by\ \bibinfo {editor} {\bibfnamefont {C.}~\bibnamefont {Enss}}}\ (\bibinfo  {publisher} {Springer Berlin Heidelberg},\ \bibinfo {address} {Berlin, Heidelberg},\ \bibinfo {year} {2005})\ pp.\ \bibinfo {pages} {151--216}\BibitemShut {NoStop}%
\bibitem [{\citenamefont {Boyd}\ \emph {et~al.}(2023)\citenamefont {Boyd}, \citenamefont {Hines}, \citenamefont {Kavner},\ and\ \citenamefont {Kim}}]{boyd2023development}%
  \BibitemOpen
  \bibfield  {author} {\bibinfo {author} {\bibfnamefont {S.~T.~P.}\ \bibnamefont {Boyd}}, \bibinfo {author} {\bibfnamefont {N.~R.}\ \bibnamefont {Hines}}, \bibinfo {author} {\bibfnamefont {A.~R.~L.}\ \bibnamefont {Kavner}}, \ and\ \bibinfo {author} {\bibfnamefont {G.-B.}\ \bibnamefont {Kim}},\ }\href {\doibase 10.1109/TASC.2023.3259334} {\bibfield  {journal} {\bibinfo  {journal} {IEEE Transactions on Applied Superconductivity}\ }\textbf {\bibinfo {volume} {33}},\ \bibinfo {pages} {1} (\bibinfo {year} {2023})}\BibitemShut {NoStop}%
\bibitem [{\citenamefont {Kim}\ \emph {et~al.}(2017)\citenamefont {Kim}, \citenamefont {Choi}, \citenamefont {Jo}, \citenamefont {Kang}, \citenamefont {Kim}, \citenamefont {Kim}, \citenamefont {Kim}, \citenamefont {Kim}, \citenamefont {Lee}, \citenamefont {Lee}, \citenamefont {Lee}, \citenamefont {Li}, \citenamefont {Oh},\ and\ \citenamefont {So}}]{amore2017novel}%
  \BibitemOpen
  \bibfield  {author} {\bibinfo {author} {\bibfnamefont {G.~B.}\ \bibnamefont {Kim}}, \bibinfo {author} {\bibfnamefont {J.~H.}\ \bibnamefont {Choi}}, \bibinfo {author} {\bibfnamefont {H.~S.}\ \bibnamefont {Jo}}, \bibinfo {author} {\bibfnamefont {C.~S.}\ \bibnamefont {Kang}}, \bibinfo {author} {\bibfnamefont {H.~L.}\ \bibnamefont {Kim}}, \bibinfo {author} {\bibfnamefont {I.}~\bibnamefont {Kim}}, \bibinfo {author} {\bibfnamefont {S.~R.}\ \bibnamefont {Kim}}, \bibinfo {author} {\bibfnamefont {Y.~H.}\ \bibnamefont {Kim}}, \bibinfo {author} {\bibfnamefont {C.}~\bibnamefont {Lee}}, \bibinfo {author} {\bibfnamefont {H.~J.}\ \bibnamefont {Lee}}, \bibinfo {author} {\bibfnamefont {M.~K.}\ \bibnamefont {Lee}}, \bibinfo {author} {\bibfnamefont {J.}~\bibnamefont {Li}}, \bibinfo {author} {\bibfnamefont {S.~Y.}\ \bibnamefont {Oh}}, \ and\ \bibinfo {author} {\bibfnamefont {J.~H.}\ \bibnamefont {So}},\ }\href {\doibase https://doi.org/10.1016/j.astropartphys.2017.02.009} {\bibfield  {journal} {\bibinfo  {journal}
  {Astroparticle Physics}\ }\textbf {\bibinfo {volume} {91}},\ \bibinfo {pages} {105} (\bibinfo {year} {2017})}\BibitemShut {NoStop}%
\bibitem [{\citenamefont {Stevenson}\ and\ \citenamefont {Keyes}(1955)}]{stevenson1955carrier}%
  \BibitemOpen
  \bibfield  {author} {\bibinfo {author} {\bibfnamefont {D.~T.}\ \bibnamefont {Stevenson}}\ and\ \bibinfo {author} {\bibfnamefont {R.~J.}\ \bibnamefont {Keyes}},\ }\href {\doibase 10.1063/1.1721958} {\bibfield  {journal} {\bibinfo  {journal} {Journal of Applied Physics}\ }\textbf {\bibinfo {volume} {26}},\ \bibinfo {pages} {190} (\bibinfo {year} {1955})},\ \Eprint {http://arxiv.org/abs/https://pubs.aip.org/aip/jap/article-pdf/26/2/190/7922020/190\_1\_online.pdf} {https://pubs.aip.org/aip/jap/article-pdf/26/2/190/7922020/190\_1\_online.pdf} \BibitemShut {NoStop}%
\bibitem [{\citenamefont {Isberg}\ \emph {et~al.}(2002)\citenamefont {Isberg}, \citenamefont {Hammersberg}, \citenamefont {Johansson}, \citenamefont {Wikstrom}, \citenamefont {Twitchen}, \citenamefont {Whitehead}, \citenamefont {Coe},\ and\ \citenamefont {Scarsbrook}}]{isberg2002high}%
  \BibitemOpen
  \bibfield  {author} {\bibinfo {author} {\bibfnamefont {J.}~\bibnamefont {Isberg}}, \bibinfo {author} {\bibfnamefont {J.}~\bibnamefont {Hammersberg}}, \bibinfo {author} {\bibfnamefont {E.}~\bibnamefont {Johansson}}, \bibinfo {author} {\bibfnamefont {T.}~\bibnamefont {Wikstrom}}, \bibinfo {author} {\bibfnamefont {D.~J.}\ \bibnamefont {Twitchen}}, \bibinfo {author} {\bibfnamefont {A.~J.}\ \bibnamefont {Whitehead}}, \bibinfo {author} {\bibfnamefont {S.~E.}\ \bibnamefont {Coe}}, \ and\ \bibinfo {author} {\bibfnamefont {G.~A.}\ \bibnamefont {Scarsbrook}},\ }\href {\doibase 10.1126/science.1074374} {\bibfield  {journal} {\bibinfo  {journal} {Science}\ }\textbf {\bibinfo {volume} {297}},\ \bibinfo {pages} {1670} (\bibinfo {year} {2002})},\ \Eprint {http://arxiv.org/abs/https://www.science.org/doi/pdf/10.1126/science.1074374} {https://www.science.org/doi/pdf/10.1126/science.1074374} \BibitemShut {NoStop}%
\bibitem [{\citenamefont {Bi}\ \emph {et~al.}(2016)\citenamefont {Bi}, \citenamefont {Hutter}, \citenamefont {Fang}, \citenamefont {Dong}, \citenamefont {Huang},\ and\ \citenamefont {Savenije}}]{bi2016charge}%
  \BibitemOpen
  \bibfield  {author} {\bibinfo {author} {\bibfnamefont {Y.}~\bibnamefont {Bi}}, \bibinfo {author} {\bibfnamefont {E.~M.}\ \bibnamefont {Hutter}}, \bibinfo {author} {\bibfnamefont {Y.}~\bibnamefont {Fang}}, \bibinfo {author} {\bibfnamefont {Q.}~\bibnamefont {Dong}}, \bibinfo {author} {\bibfnamefont {J.}~\bibnamefont {Huang}}, \ and\ \bibinfo {author} {\bibfnamefont {T.~J.}\ \bibnamefont {Savenije}},\ }\href {\doibase 10.1021/acs.jpclett.6b00269} {\bibfield  {journal} {\bibinfo  {journal} {The journal of physical chemistry letters}\ }\textbf {\bibinfo {volume} {7}},\ \bibinfo {pages} {923} (\bibinfo {year} {2016})}\BibitemShut {NoStop}%
\bibitem [{\citenamefont {Zhang}\ \emph {et~al.}(2017)\citenamefont {Zhang}, \citenamefont {Yang}, \citenamefont {Mao}, \citenamefont {Yang}, \citenamefont {Jiang}, \citenamefont {Li}, \citenamefont {Xiong}, \citenamefont {Yang}, \citenamefont {He}, \citenamefont {Deng} \emph {et~al.}}]{zhang2017perovskite}%
  \BibitemOpen
  \bibfield  {author} {\bibinfo {author} {\bibfnamefont {F.}~\bibnamefont {Zhang}}, \bibinfo {author} {\bibfnamefont {B.}~\bibnamefont {Yang}}, \bibinfo {author} {\bibfnamefont {X.}~\bibnamefont {Mao}}, \bibinfo {author} {\bibfnamefont {R.}~\bibnamefont {Yang}}, \bibinfo {author} {\bibfnamefont {L.}~\bibnamefont {Jiang}}, \bibinfo {author} {\bibfnamefont {Y.}~\bibnamefont {Li}}, \bibinfo {author} {\bibfnamefont {J.}~\bibnamefont {Xiong}}, \bibinfo {author} {\bibfnamefont {Y.}~\bibnamefont {Yang}}, \bibinfo {author} {\bibfnamefont {R.}~\bibnamefont {He}}, \bibinfo {author} {\bibfnamefont {W.}~\bibnamefont {Deng}},  \emph {et~al.},\ }\href {\doibase 10.1021/acsami.7b01696} {\bibfield  {journal} {\bibinfo  {journal} {ACS Applied Materials \& Interfaces}\ }\textbf {\bibinfo {volume} {9}},\ \bibinfo {pages} {14827} (\bibinfo {year} {2017})}\BibitemShut {NoStop}%
\bibitem [{\citenamefont {Gradwohl}\ \emph {et~al.}(2021)\citenamefont {Gradwohl}, \citenamefont {Juda},\ and\ \citenamefont {{Radhakrishnan Sumathi}}}]{gradwohl2021carrier}%
  \BibitemOpen
  \bibfield  {author} {\bibinfo {author} {\bibfnamefont {K.-P.}\ \bibnamefont {Gradwohl}}, \bibinfo {author} {\bibfnamefont {U.}~\bibnamefont {Juda}}, \ and\ \bibinfo {author} {\bibfnamefont {R.}~\bibnamefont {{Radhakrishnan Sumathi}}},\ }\href {\doibase https://doi.org/10.1016/j.jcrysgro.2021.126285} {\bibfield  {journal} {\bibinfo  {journal} {Journal of Crystal Growth}\ }\textbf {\bibinfo {volume} {573}},\ \bibinfo {pages} {126285} (\bibinfo {year} {2021})}\BibitemShut {NoStop}%
\bibitem [{\citenamefont {de~Boer}\ \emph {et~al.}(2007)\citenamefont {de~Boer}, \citenamefont {Bol}, \citenamefont {Furgeri}, \citenamefont {Müller}, \citenamefont {Sander}, \citenamefont {Berdermann}, \citenamefont {Pomorski},\ and\ \citenamefont {Huhtinen}}]{deboer2007radiation}%
  \BibitemOpen
  \bibfield  {author} {\bibinfo {author} {\bibfnamefont {W.}~\bibnamefont {de~Boer}}, \bibinfo {author} {\bibfnamefont {J.}~\bibnamefont {Bol}}, \bibinfo {author} {\bibfnamefont {A.}~\bibnamefont {Furgeri}}, \bibinfo {author} {\bibfnamefont {S.}~\bibnamefont {Müller}}, \bibinfo {author} {\bibfnamefont {C.}~\bibnamefont {Sander}}, \bibinfo {author} {\bibfnamefont {E.}~\bibnamefont {Berdermann}}, \bibinfo {author} {\bibfnamefont {M.}~\bibnamefont {Pomorski}}, \ and\ \bibinfo {author} {\bibfnamefont {M.}~\bibnamefont {Huhtinen}},\ }\href {\doibase https://doi.org/10.1002/pssa.200776327} {\bibfield  {journal} {\bibinfo  {journal} {physica status solidi (a)}\ }\textbf {\bibinfo {volume} {204}},\ \bibinfo {pages} {3004} (\bibinfo {year} {2007})},\ \Eprint {http://arxiv.org/abs/https://onlinelibrary.wiley.com/doi/pdf/10.1002/pssa.200776327} {https://onlinelibrary.wiley.com/doi/pdf/10.1002/pssa.200776327} \BibitemShut {NoStop}%
\bibitem [{\citenamefont {Koz{\'a}k}\ \emph {et~al.}(2013)\citenamefont {Koz{\'a}k}, \citenamefont {Troj{\'a}nek},\ and\ \citenamefont {Mal{\`y}}}]{kozak2013optical}%
  \BibitemOpen
  \bibfield  {author} {\bibinfo {author} {\bibfnamefont {M.}~\bibnamefont {Koz{\'a}k}}, \bibinfo {author} {\bibfnamefont {F.}~\bibnamefont {Troj{\'a}nek}}, \ and\ \bibinfo {author} {\bibfnamefont {P.}~\bibnamefont {Mal{\`y}}},\ }\href@noop {} {\bibfield  {journal} {\bibinfo  {journal} {physica status solidi (a)}\ }\textbf {\bibinfo {volume} {210}},\ \bibinfo {pages} {2008} (\bibinfo {year} {2013})}\BibitemShut {NoStop}%
\bibitem [{\citenamefont {Shimano}\ \emph {et~al.}(2002)\citenamefont {Shimano}, \citenamefont {Nagai}, \citenamefont {Horiuch},\ and\ \citenamefont {Kuwata-Gonokami}}]{shimano2002formation}%
  \BibitemOpen
  \bibfield  {author} {\bibinfo {author} {\bibfnamefont {R.}~\bibnamefont {Shimano}}, \bibinfo {author} {\bibfnamefont {M.}~\bibnamefont {Nagai}}, \bibinfo {author} {\bibfnamefont {K.}~\bibnamefont {Horiuch}}, \ and\ \bibinfo {author} {\bibfnamefont {M.}~\bibnamefont {Kuwata-Gonokami}},\ }\href@noop {} {\bibfield  {journal} {\bibinfo  {journal} {Physical review letters}\ }\textbf {\bibinfo {volume} {88}},\ \bibinfo {pages} {057404} (\bibinfo {year} {2002})}\BibitemShut {NoStop}%
\bibitem [{\citenamefont {Akimoto}\ \emph {et~al.}(2014)\citenamefont {Akimoto}, \citenamefont {Handa}, \citenamefont {Fukai},\ and\ \citenamefont {Naka}}]{akimoto2014high}%
  \BibitemOpen
  \bibfield  {author} {\bibinfo {author} {\bibfnamefont {I.}~\bibnamefont {Akimoto}}, \bibinfo {author} {\bibfnamefont {Y.}~\bibnamefont {Handa}}, \bibinfo {author} {\bibfnamefont {K.}~\bibnamefont {Fukai}}, \ and\ \bibinfo {author} {\bibfnamefont {N.}~\bibnamefont {Naka}},\ }\href@noop {} {\bibfield  {journal} {\bibinfo  {journal} {Applied Physics Letters}\ }\textbf {\bibinfo {volume} {105}} (\bibinfo {year} {2014})}\BibitemShut {NoStop}%
\bibitem [{\citenamefont {Ziaja}\ \emph {et~al.}(2005)\citenamefont {Ziaja}, \citenamefont {London},\ and\ \citenamefont {Hajdu}}]{ziaja2005cascades}%
  \BibitemOpen
  \bibfield  {author} {\bibinfo {author} {\bibfnamefont {B.}~\bibnamefont {Ziaja}}, \bibinfo {author} {\bibfnamefont {R.~A.}\ \bibnamefont {London}}, \ and\ \bibinfo {author} {\bibfnamefont {J.}~\bibnamefont {Hajdu}},\ }\href {\doibase 10.1063/1.1853494} {\bibfield  {journal} {\bibinfo  {journal} {Journal of Applied Physics}\ }\textbf {\bibinfo {volume} {97}},\ \bibinfo {pages} {064905} (\bibinfo {year} {2005})},\ \Eprint {http://arxiv.org/abs/https://pubs.aip.org/aip/jap/article-pdf/doi/10.1063/1.1853494/14943355/064905\_1\_online.pdf} {https://pubs.aip.org/aip/jap/article-pdf/doi/10.1063/1.1853494/14943355/064905\_1\_online.pdf} \BibitemShut {NoStop}%
\bibitem [{\citenamefont {Kim}(2020)}]{kim2020self}%
  \BibitemOpen
  \bibfield  {author} {\bibinfo {author} {\bibfnamefont {G.-B.}\ \bibnamefont {Kim}},\ }\href@noop {} {\bibfield  {journal} {\bibinfo  {journal} {Journal of Low Temperature Physics}\ }\textbf {\bibinfo {volume} {199}},\ \bibinfo {pages} {1004} (\bibinfo {year} {2020})}\BibitemShut {NoStop}%
\bibitem [{\citenamefont {Zhang}\ \emph {et~al.}(2015)\citenamefont {Zhang}, \citenamefont {Lin}, \citenamefont {Mikhailik},\ and\ \citenamefont {Kraus}}]{amore2015scintillation}%
  \BibitemOpen
  \bibfield  {author} {\bibinfo {author} {\bibfnamefont {X.}~\bibnamefont {Zhang}}, \bibinfo {author} {\bibfnamefont {J.}~\bibnamefont {Lin}}, \bibinfo {author} {\bibfnamefont {V.~B.}\ \bibnamefont {Mikhailik}}, \ and\ \bibinfo {author} {\bibfnamefont {H.}~\bibnamefont {Kraus}},\ }\href {\doibase 10.1063/1.4922875} {\bibfield  {journal} {\bibinfo  {journal} {Applied Physics Letters}\ }\textbf {\bibinfo {volume} {106}},\ \bibinfo {pages} {241904} (\bibinfo {year} {2015})},\ \Eprint {http://arxiv.org/abs/https://pubs.aip.org/aip/apl/article-pdf/doi/10.1063/1.4922875/13280637/241904\_1\_online.pdf} {https://pubs.aip.org/aip/apl/article-pdf/doi/10.1063/1.4922875/13280637/241904\_1\_online.pdf} \BibitemShut {NoStop}%
\bibitem [{\citenamefont {Kim}(2023)}]{magnetodm2023}%
  \BibitemOpen
  \bibfield  {author} {\bibinfo {author} {\bibfnamefont {G.-B.}\ \bibnamefont {Kim}},\ }\href {\doibase 10.2172/1972903} {\  (\bibinfo {year} {2023}),\ 10.2172/1972903}\BibitemShut {NoStop}%
\bibitem [{\citenamefont {Sun}\ \emph {et~al.}(1994)\citenamefont {Sun}, \citenamefont {Vall\'ee}, \citenamefont {Acioli}, \citenamefont {Ippen},\ and\ \citenamefont {Fujimoto}}]{sun1994femtosecond}%
  \BibitemOpen
  \bibfield  {author} {\bibinfo {author} {\bibfnamefont {C.-K.}\ \bibnamefont {Sun}}, \bibinfo {author} {\bibfnamefont {F.}~\bibnamefont {Vall\'ee}}, \bibinfo {author} {\bibfnamefont {L.~H.}\ \bibnamefont {Acioli}}, \bibinfo {author} {\bibfnamefont {E.~P.}\ \bibnamefont {Ippen}}, \ and\ \bibinfo {author} {\bibfnamefont {J.~G.}\ \bibnamefont {Fujimoto}},\ }\href {\doibase 10.1103/PhysRevB.50.15337} {\bibfield  {journal} {\bibinfo  {journal} {Phys. Rev. B}\ }\textbf {\bibinfo {volume} {50}},\ \bibinfo {pages} {15337} (\bibinfo {year} {1994})}\BibitemShut {NoStop}%
\bibitem [{\citenamefont {Fann}\ \emph {et~al.}(1992)\citenamefont {Fann}, \citenamefont {Storz}, \citenamefont {Tom},\ and\ \citenamefont {Bokor}}]{fann1992electron}%
  \BibitemOpen
  \bibfield  {author} {\bibinfo {author} {\bibfnamefont {W.~S.}\ \bibnamefont {Fann}}, \bibinfo {author} {\bibfnamefont {R.}~\bibnamefont {Storz}}, \bibinfo {author} {\bibfnamefont {H.~W.~K.}\ \bibnamefont {Tom}}, \ and\ \bibinfo {author} {\bibfnamefont {J.}~\bibnamefont {Bokor}},\ }\href {\doibase 10.1103/PhysRevB.46.13592} {\bibfield  {journal} {\bibinfo  {journal} {Phys. Rev. B}\ }\textbf {\bibinfo {volume} {46}},\ \bibinfo {pages} {13592} (\bibinfo {year} {1992})}\BibitemShut {NoStop}%
\bibitem [{\citenamefont {Pr{\"o}bst}\ \emph {et~al.}(1995)\citenamefont {Pr{\"o}bst}, \citenamefont {Frank}, \citenamefont {Cooper}, \citenamefont {Colling}, \citenamefont {Dummer}, \citenamefont {Ferger}, \citenamefont {Forster}, \citenamefont {Nucciotti}, \citenamefont {Seidel},\ and\ \citenamefont {Stodolsky}}]{probst1995model}%
  \BibitemOpen
  \bibfield  {author} {\bibinfo {author} {\bibfnamefont {F.}~\bibnamefont {Pr{\"o}bst}}, \bibinfo {author} {\bibfnamefont {M.}~\bibnamefont {Frank}}, \bibinfo {author} {\bibfnamefont {S.}~\bibnamefont {Cooper}}, \bibinfo {author} {\bibfnamefont {P.}~\bibnamefont {Colling}}, \bibinfo {author} {\bibfnamefont {D.}~\bibnamefont {Dummer}}, \bibinfo {author} {\bibfnamefont {P.}~\bibnamefont {Ferger}}, \bibinfo {author} {\bibfnamefont {G.}~\bibnamefont {Forster}}, \bibinfo {author} {\bibfnamefont {A.}~\bibnamefont {Nucciotti}}, \bibinfo {author} {\bibfnamefont {W.}~\bibnamefont {Seidel}}, \ and\ \bibinfo {author} {\bibfnamefont {L.}~\bibnamefont {Stodolsky}},\ }\href {\doibase 10.1007/BF00753837} {\bibfield  {journal} {\bibinfo  {journal} {Journal of low temperature physics}\ }\textbf {\bibinfo {volume} {100}},\ \bibinfo {pages} {69} (\bibinfo {year} {1995})}\BibitemShut {NoStop}%
\bibitem [{\citenamefont {{II-VI Incorporated}}()}]{iivi2023}%
  \BibitemOpen
  \bibfield  {author} {\bibinfo {author} {\bibnamefont {{II-VI Incorporated}}},\ }\href@noop {} {}\bibinfo {note} {\url{https://ii-vi.com/}}\BibitemShut {NoStop}%
\bibitem [{\citenamefont {{Element Six}}()}]{e6}%
  \BibitemOpen
  \bibfield  {author} {\bibinfo {author} {\bibnamefont {{Element Six}}},\ }\href@noop {} {}\bibinfo {note} {\url{https://www.e6.com/}}\BibitemShut {NoStop}%
\bibitem [{\citenamefont {{National Instruments}}()}]{nationalinstrument}%
  \BibitemOpen
  \bibfield  {author} {\bibinfo {author} {\bibnamefont {{National Instruments}}},\ }\href@noop {} {}\bibinfo {note} {\url{https://www.ni.com/}}\BibitemShut {NoStop}%
\bibitem [{\citenamefont {{Stanford Research Systems}}()}]{srs}%
  \BibitemOpen
  \bibfield  {author} {\bibinfo {author} {\bibnamefont {{Stanford Research Systems}}},\ }\href@noop {} {}\bibinfo {note} {\url{https://www.thinksrs.com/}}\BibitemShut {NoStop}%
\bibitem [{\citenamefont {Jordanov}\ \emph {et~al.}(1994)\citenamefont {Jordanov}, \citenamefont {Knoll}, \citenamefont {Huber},\ and\ \citenamefont {Pantazis}}]{jordanov1994digital}%
  \BibitemOpen
  \bibfield  {author} {\bibinfo {author} {\bibfnamefont {V.~T.}\ \bibnamefont {Jordanov}}, \bibinfo {author} {\bibfnamefont {G.~F.}\ \bibnamefont {Knoll}}, \bibinfo {author} {\bibfnamefont {A.~C.}\ \bibnamefont {Huber}}, \ and\ \bibinfo {author} {\bibfnamefont {J.~A.}\ \bibnamefont {Pantazis}},\ }\href {\doibase https://doi.org/10.1016/0168-9002(94)91652-7} {\bibfield  {journal} {\bibinfo  {journal} {Nuclear Instruments and Methods in Physics Research Section A: Accelerators, Spectrometers, Detectors and Associated Equipment}\ }\textbf {\bibinfo {volume} {353}},\ \bibinfo {pages} {261} (\bibinfo {year} {1994})}\BibitemShut {NoStop}%
\bibitem [{\citenamefont {Liu}\ \emph {et~al.}(2017{\natexlab{b}})\citenamefont {Liu}, \citenamefont {Ouyang}, \citenamefont {Zhang}, \citenamefont {Jin},\ and\ \citenamefont {Su}}]{liu2017diamond}%
  \BibitemOpen
  \bibfield  {author} {\bibinfo {author} {\bibfnamefont {L.}~\bibnamefont {Liu}}, \bibinfo {author} {\bibfnamefont {X.}~\bibnamefont {Ouyang}}, \bibinfo {author} {\bibfnamefont {J.}~\bibnamefont {Zhang}}, \bibinfo {author} {\bibfnamefont {P.}~\bibnamefont {Jin}}, \ and\ \bibinfo {author} {\bibfnamefont {C.}~\bibnamefont {Su}},\ }\href {\doibase https://doi.org/10.1016/j.diamond.2016.10.002} {\bibfield  {journal} {\bibinfo  {journal} {Diamond and Related Materials}\ }\textbf {\bibinfo {volume} {73}},\ \bibinfo {pages} {248} (\bibinfo {year} {2017}{\natexlab{b}})},\ \bibinfo {note} {10th International Conference on New Diamond and Nano Carbons – NDNC 2016}\BibitemShut {NoStop}%
\bibitem [{\citenamefont {Gallin-Martel}\ \emph {et~al.}(2021)\citenamefont {Gallin-Martel}, \citenamefont {Kim}, \citenamefont {Abbassi}, \citenamefont {Bes}, \citenamefont {Boiano}, \citenamefont {Brambilla}, \citenamefont {Collot}, \citenamefont {Colombi}, \citenamefont {Crozes}, \citenamefont {Curtoni} \emph {et~al.}}]{gallin2021characterization}%
  \BibitemOpen
  \bibfield  {author} {\bibinfo {author} {\bibfnamefont {M.~L.}\ \bibnamefont {Gallin-Martel}}, \bibinfo {author} {\bibfnamefont {Y.~H.}\ \bibnamefont {Kim}}, \bibinfo {author} {\bibfnamefont {L.}~\bibnamefont {Abbassi}}, \bibinfo {author} {\bibfnamefont {A.}~\bibnamefont {Bes}}, \bibinfo {author} {\bibfnamefont {C.}~\bibnamefont {Boiano}}, \bibinfo {author} {\bibfnamefont {S.}~\bibnamefont {Brambilla}}, \bibinfo {author} {\bibfnamefont {J.}~\bibnamefont {Collot}}, \bibinfo {author} {\bibfnamefont {G.}~\bibnamefont {Colombi}}, \bibinfo {author} {\bibfnamefont {T.}~\bibnamefont {Crozes}}, \bibinfo {author} {\bibfnamefont {S.}~\bibnamefont {Curtoni}},  \emph {et~al.},\ }\href {\doibase 10.3389/fphy.2021.732730} {\bibfield  {journal} {\bibinfo  {journal} {Frontiers in Physics}\ }\textbf {\bibinfo {volume} {9}},\ \bibinfo {pages} {732730} (\bibinfo {year} {2021})}\BibitemShut {NoStop}%
\bibitem [{\citenamefont {Fischler}\ and\ \citenamefont {Bolles}(1981)}]{ransac1981}%
  \BibitemOpen
  \bibfield  {author} {\bibinfo {author} {\bibfnamefont {M.~A.}\ \bibnamefont {Fischler}}\ and\ \bibinfo {author} {\bibfnamefont {R.~C.}\ \bibnamefont {Bolles}},\ }\href {\doibase 10.1145/358669.358692} {\bibfield  {journal} {\bibinfo  {journal} {Commun. ACM}\ }\textbf {\bibinfo {volume} {24}},\ \bibinfo {pages} {381–395} (\bibinfo {year} {1981})}\BibitemShut {NoStop}%
\bibitem [{\citenamefont {Adari}\ \emph {et~al.}(2022)\citenamefont {Adari}, \citenamefont {Aguilar-Arevalo}, \citenamefont {Amidei}, \citenamefont {Angloher}, \citenamefont {Armengaud}, \citenamefont {Augier}, \citenamefont {Balogh}, \citenamefont {Banik}, \citenamefont {Baxter}, \citenamefont {Beaufort} \emph {et~al.}}]{excess2022}%
  \BibitemOpen
  \bibfield  {author} {\bibinfo {author} {\bibfnamefont {P.}~\bibnamefont {Adari}}, \bibinfo {author} {\bibfnamefont {A.}~\bibnamefont {Aguilar-Arevalo}}, \bibinfo {author} {\bibfnamefont {D.}~\bibnamefont {Amidei}}, \bibinfo {author} {\bibfnamefont {G.}~\bibnamefont {Angloher}}, \bibinfo {author} {\bibfnamefont {E.}~\bibnamefont {Armengaud}}, \bibinfo {author} {\bibfnamefont {C.}~\bibnamefont {Augier}}, \bibinfo {author} {\bibfnamefont {L.}~\bibnamefont {Balogh}}, \bibinfo {author} {\bibfnamefont {S.}~\bibnamefont {Banik}}, \bibinfo {author} {\bibfnamefont {D.}~\bibnamefont {Baxter}}, \bibinfo {author} {\bibfnamefont {C.}~\bibnamefont {Beaufort}},  \emph {et~al.},\ }\href {\doibase 10.21468/SciPostPhysProc.9.001} {\bibfield  {journal} {\bibinfo  {journal} {SciPost Phys. Proc.}\ ,\ \bibinfo {pages} {001}} (\bibinfo {year} {2022})}\BibitemShut {NoStop}%
\end{thebibliography}%

\end{document}